\documentclass[fleqn,usenatbib]{mnras}

\usepackage{newtxtext,newtxmath}

\usepackage[T1]{fontenc}
\usepackage{ae,aecompl}

\usepackage{graphicx}	
\usepackage{amsmath}	
\usepackage{amssymb}	
\newcommand{\Teff}{$T_\mathrm{eff}$}
\newcommand{\Logg}{$\log g$}
\newcommand{\Feh}{$\mathrm{[Fe/H]}$}
\newcommand{\Meh}{$\mathrm{[M/H]}$}
\newcommand{\Cfe}{$\mathrm{[C/Fe]}$}
\newcommand{\CO}{$\mathrm{[C/O]}$}

\newcommand{\TC}{\textit{The Cannon}}
\newcommand{\kms}{\,km\,s$^{-1}$}

\title[GALAH survey: CEMP candidates]{The GALAH survey: a catalogue of carbon-enhanced stars and CEMP candidates}

\author[K. \v{C}otar et al.]{
Klemen~\v{C}otar,$^{1}$\thanks{Contact e-mail: \href{mailto:klemen.cotar@fmf.uni-lj.si}{klemen.cotar@fmf.uni-lj.si}}
Toma\v{z}~Zwitter,$^{1}$ Janez Kos,$^{1,2}$ Ulisse Munari,$^{3}$ Sarah L. Martell,$^{4,5}$\newauthor
Martin Asplund,$^{5,6}$ Joss Bland-Hawthorn,$^{2,5}$ Sven Buder,$^{7}$ Gayandhi M. De Silva,$^{8}$\newauthor 
Kenneth C. Freeman,$^{6}$ Sanjib Sharma,$^{2}$ Borja Anguiano,$^{9}$ Daniela Carollo,$^{10}$\newauthor Jonathan Horner,$^{11}$ Geraint F. Lewis,$^{2}$  David M. Nataf,$^{12}$ Thomas Nordlander,$^{5,6}$\newauthor
Denis Stello,$^{2,4,5,13}$ Yuan-Sen Ting,$^{14,15,16}$ Chris Tinney,$^{4}$ Gregor Traven,$^{1,17}$\newauthor
Rob A. Wittenmyer,$^{11}$ and~the~GALAH~collaboration
\\
\\
$^{1}$ Faculty of Mathematics and Physics, University of Ljubljana, Jadranska 19, 1000 Ljubljana, Slovenia\\
$^{2}$ Sydney Institute for Astronomy, School of Physics, A28, The
University of Sydney, NSW 2006, Australia\\
$^{3}$ INAF, Osservatorio Astronomico di Padova, Sede di Asiago, I-36012 Asiago (VI), Italy\\
$^{4}$ School of Physics, UNSW, Sydney, NSW 2052, Australia\\ 
$^{5}$ ARC Centre of Excellence for All-Sky Astrophysics in Three Dimensions (ASTRO 3D), Australia\\
$^{6}$ Research School of Astronomy \& Astrophysics, Australian National University, ACT 2611, Australia\\
$^{7}$ Max Planck Institute for Astronomy (MPIA), Koenigstuhl 17, D-69117 Heidelberg, Germany\\
$^{8}$ AAO-MQ, Macquarie University, Sydney, NSW 2109, Australia \\
$^{9}$ Department of Astronomy, University of Virginia, Charlottesville, VA 22904-4325, USA \\
$^{10}$ INAF, Astrophysical Observatory of Turin, Torino, Italy\\
$^{11}$ University of Southern Queensland, Toowoomba, Queensland 4350, Australia \\
$^{12}$ Department of Physics and Astronomy, The Johns Hopkins University, Baltimore, MD 21218, USA\\
$^{13}$ Stellar Astrophysics Centre, Department of Physics and Astronomy, Aarhus University, Denmark\\
$^{14}$ Institute for Advanced Study, Princeton, NJ 08540, USA\\
$^{15}$ Observatories of the Carnegie Institution of Washington, 813 Santa Barbara Street, Pasadena, CA 91101, USA\\
$^{16}$ Department of Astrophysical Sciences, Princeton University, Princeton, NJ 08544, USA \\
$^{17}$ Lund Observatory, Department of Astronomy and Theoretical Physics, Box 43, SE-221 00 Lund, Sweden
}

\date{Accepted XXX. Received YYY; in original form ZZZ}

\pubyear{2018}

\begin{document}
\label{firstpage}
\pagerange{\pageref{firstpage}--\pageref{lastpage}}
\maketitle

\begin{abstract}
Swan bands  -- characteristic molecular absorption features of the C$_2$ molecule -- are a spectroscopic signature of carbon-enhanced stars. They can also be used to identify carbon-enhanced metal-poor (CEMP) stars. The GALAH (GALactic Archaeology with Hermes) is a magnitude-limited survey of stars producing high-resolution, high signal-to-noise spectra. We used 627,708 GALAH spectra to search for carbon-enhanced stars with a supervised and unsupervised classification algorithm, relying on the imprint of the Swan bands. We identified 918 carbon-enhanced stars, including 12 already described in the literature. An unbiased selection function of the GALAH survey allows us to perform a population study of carbon-enhanced stars. Most of them are giants, out of which we find 28 CEMP candidates. A large fraction of our carbon-enhanced stars with repeated observations show variation in radial velocity, hinting that there is a large fraction of variables among them. 32 of the detected stars also show strong Lithium enhancement in their spectra.
\end{abstract}
\begin{keywords}
methods: data analysis -- stars: carbon -- stars: abundances -- catalogues
\end{keywords}

\section{Introduction}
Chemically peculiar stars whose spectra are dominated by carbon molecular bands were first identified by \citet{1869AN.....73..129S}. Their spectra are characterised by enhanced carbon absorption bands of CH, CN, SiC$_2$, and C$_{2}$ molecules, also known as Swan bands. Possible sources of enhancement are dredge-up events in evolved stars \citep{1983ApJ...275L..65I}, enrichment by carbon-rich stellar winds from a pulsating asymptotic giant branch (AGB) star, which settles on a main sequence companion \citep{1995MNRAS.277.1443H}, or it can be the result of a primordial enrichment \citep{2016ApJ...833...20Y}. Historically, high latitude carbon stars, presumed to be giants, were used as probes to measure the Galactic rotation curve \citep{2013Ap.....56...68B}, velocity dispersion in the Galactic halo \citep{1991AJ....101.2220B}, and to trace the gravitational potential of the Galaxy.  

Because of their strong spectral features, the most prominent candidates can easily be identified from large photometric surveys \citep{2002AJ....124.1651M, 2004AJ....127.2838D}. Specific photometric systems \citep{1960MNRAS.120..287G, 1968AJ.....73..313M, 1970A&AS....1..199H} were defined in the past to discover and further classify stars with enhanced carbon features in their spectra. Specifics of those systems were catalogued, compared, and homogenised by \citet{2000A&AS..147..361M} and \citet{2003A&A...401..781F}.

Other useful data come from low-resolution spectroscopic surveys, whose classification identified from a few hundred to a few thousand of those objects \citep{2001A&A...375..366C, 2013ApJ...765...12G, 2013AJ....146..132L, 2016ApJS..226....1J, 2018ApJS..234...31L}. High-resolution spectroscopy is required to search for candidates with less pronounced molecular absorption features or to determine their stellar chemical composition. Multiple studies have been carried out to determine accurate abundances of metal-poor stars \citep{1997ApJ...488..350N, 2002ApJ...567.1166A, 2004A&A...416.1117C, 2005ESASP.560..433B, 2006AJ....132..137C, 2007ApJ...655..492A, 2007ApJ...670..774N, 2011ApJ...742...54H, 2013ApJ...762...26Y, 2014AJ....147..136R, 2015ApJ...807..173H, 2015ApJ...807..171J}. Such detailed abundance information is especially important for the analysis and classification of chemically peculiar objects \citep{2013ApJ...778...56C}.

Today, the most sought after, of all carbon-enhanced stars, are the carbon-enhanced metal-poor (CEMP) ones whose fraction, among metal-poor stars, increases with decreasing metallicity \Meh\ \citep{1992AJ....103.1987B, 1997ApJ...488..350N, 1999ASPC..165..264R, 2005ApJ...633L.109C, 2005ApJ...625..825L, 2005AJ....130.2804R, 2006ApJ...652.1585F, 2007PhDT........22M, 2012ApJ...744..195C, 2013AJ....146..132L, 2013ApJ...762...27Y,  2014ApJ...797...21P, 2018ApJ...861..146Y}. Amongst these, those near the main-sequence turn-off are expected to be of particular importance, as they may have accreted enough material from their AGB companion to produce an observable change in their atmospheric chemical composition \citep{2004ApJ...611..476S, 2014MNRAS.441.1217S, 2015ApJ...807..173H}. The accreted material could provide insight into the production efficiency of neutron capture elements in AGB stars \citep{2007ApJ...655..492A}. Multiple studies show that a peculiar observed abundance pattern and carbon enrichment in a certain type of CEMP stars could be explained by the supernova explosions of first-generation stars that enriched the interstellar medium \citep{2003Natur.422..871U, 2005ApJ...619..427U, 2014ApJ...785...98T, 2018MNRAS.tmp.2127B}. The exact origin and underlying physical processes governing multiple classes of CEMP stars are not yet fully understood and are a topic of ongoing research \citep{2014ApJ...788..180C, 2016ApJ...833...20Y, 2018MNRAS.475.4781C}. Classification into multiple sub-classes is performed using the abundance information of neutron-capture elements \citep{2005ARA&A..43..531B, 2013A&A...552A.107S, 2015ApJ...814..121H, 2016ApJ...833...20Y} that are thought to originate from different astrophysical phenomena responsible for the synthesis of those elements.

In this work, we propose a novel approach for the classification of carbon-enhanced stars using high-resolution stellar spectra covering parts of the visible domain. The goal is to identify a representative sample of carbon-enhanced stars, which can be used as an input to population studies. The paper is organised as follows; we start with a brief discussion of our spectroscopic observations and their reduction (Section \ref{sec:data}), which is followed by the description of the used algorithms for the detection of carbon-enhanced stars in Section \ref{sec:classification}. Properties of the classified objects are investigated in Section \ref{sec:analysis}, CEMP candidates are a focus of Section \ref{sec:cemp}, with Section \ref{sec:asiago} describing a follow-up study for one of them. Final remarks are given in Section \ref{sec:summary}.

\section{Data}
\label{sec:data}
The analysed set of stellar spectra was acquired by the High Efficiency and Resolution Multi-Element Spectrograph (HERMES), a fibre-fed multi-object spectrograph on the $3.9$ m Anglo-Australian Telescope (AAT) of the Australian Astronomical Observatory. The spectograph \citep{2010SPIE.7735E..09B, 2015JATIS...1c5002S} can simultaneously record spectra from up to 392 fibres distributed over a $2^\circ$ field of the night sky, with an additional 8 fibres used for the telescope guiding. The spectrograph has a resolving power of R~$\sim28,000$ and consists of four spectral arms centred at 4800, 5761, 6610, and 7740~\AA, together covering approximately 1000~\AA, including the H$\alpha$, and H$\beta$ lines. Three dichroic beam splitters are used to separate incoming light into four separated colour beams that are analysed independently. The spectrograph can typically achieve a signal to noise ratio (SNR) $\sim100$ per resolution element at magnitude V=14 in the red arm during a 1-hour long exposure. 

Spectra used in this study have been taken from multiple different observing programmes using this spectrograph: the GALactic Archaeology with HERMES (GALAH) pilot survey \citep{2018MNRAS.tmp..504D}, the main GALAH survey \citep{2015MNRAS.449.2604D}, the K2-HERMES survey \citep{2016AAS...22713816W}, and the TESS-HERMES survey \citep{2018MNRAS.473.2004S}. Most of those observing programmes exclude fields close to the Galactic plane (due to problems with high stellar density and Galactic extinction) or far away from it (not enough suitable targets to use all fibres), employ subtle different selection functions (position, limiting magnitude, crowding requirement, and photometric quality), but share the same observing procedures, reduction, and analysis pipeline \citep[internal version 5.3,][]{2017MNRAS.464.1259K}. All programmes, except the pilot survey, are magnitude-limited, with no colour cuts. This leads to an unbiased sample of stars distributed mostly across the southern sky that can be used for different population studies. Additionally, all objects from different observing programmes are analysed with the same procedure named \TC\ \citep[internal version 180325,][]{2015ApJ...808...16N, buder2018}, so their stellar parameters are determined in a consistent manner and are hence comparable across the different programmes. \TC\ algorithm employs a data-driven interpolation approach trained on a set of high-quality benchmark stars spanning the majority of the stellar parameter space \citep[for details, see][]{buder2018}. 

The spectrum synthesis code Spectroscopy Made Easy \citep[SME, ][]{1996A&AS..118..595V, 2017A&A...597A..16P} was run on the spectra in the training set to determine their stellar parameters and atmospheric chemical abundances. These values were used to train \TC\ model, which was then run on every observed spectrum to determine its stellar parameters. Every determined parameter is accompanied by a quality flag identifying its usefulness and possible problems with grid interpolation. Therefore, we can easily remove parameters that were determined for stellar spectra far away from the training set or with possible problems in any of the HERMES spectral arms. An interpolation method proved beneficial in determining stellar physical parameters and up to 23 abundances (extendable up to 31 in future versions) for every analysed star, as described in \citet{buder2018}. The parameter \Feh\ determined by \TC\ refers to an iron abundance and not overall metallicity \Meh\ as often used in the literature. As both notations are used interchangeably in the literature to describe stellar metallicity, care must be taken when comparing those two measurements.

\TC\ approach has an advantage of treating stars with a peculiar composition, such as carbon-enhanced stars, in a consistent manner with common objects, thus avoiding arbitrary jumps or offsets that would depend on a degree of peculiarity or correctness of physics of the underlying model. As a data-driven approach, it only projects the learned training set onto the whole survey based on quadratic relations to stellar parameters.

Our data set consists of 627,708 successfully reduced spectra of 576,229 stars observed between November 2013 and February 2018.

\section{Detection procedure}
\label{sec:classification}
To search for carbon-enhanced stars in the GALAH data set we focused on spectral features that can be clearly distinguished and are known markers of carbon enhancement. Instead of using one very weak atomic carbon absorption line (at 6587.61~\AA), used by \TC\ to determine \Cfe\ abundance, we focused on a region between 4718 and 4760~\AA\ observed in the blue arm that covers 4718--4903~\AA\ in its rest-frame. In this range we can, depending on the radial velocity of the star, observe at least four Swan band features \citep{1927RSPTA.226..157J} with their band heads located at approximately 4715, 4722, 4737, and 4745~\AA. 

\begin{figure*}
	\centering
	\includegraphics[width=\textwidth]{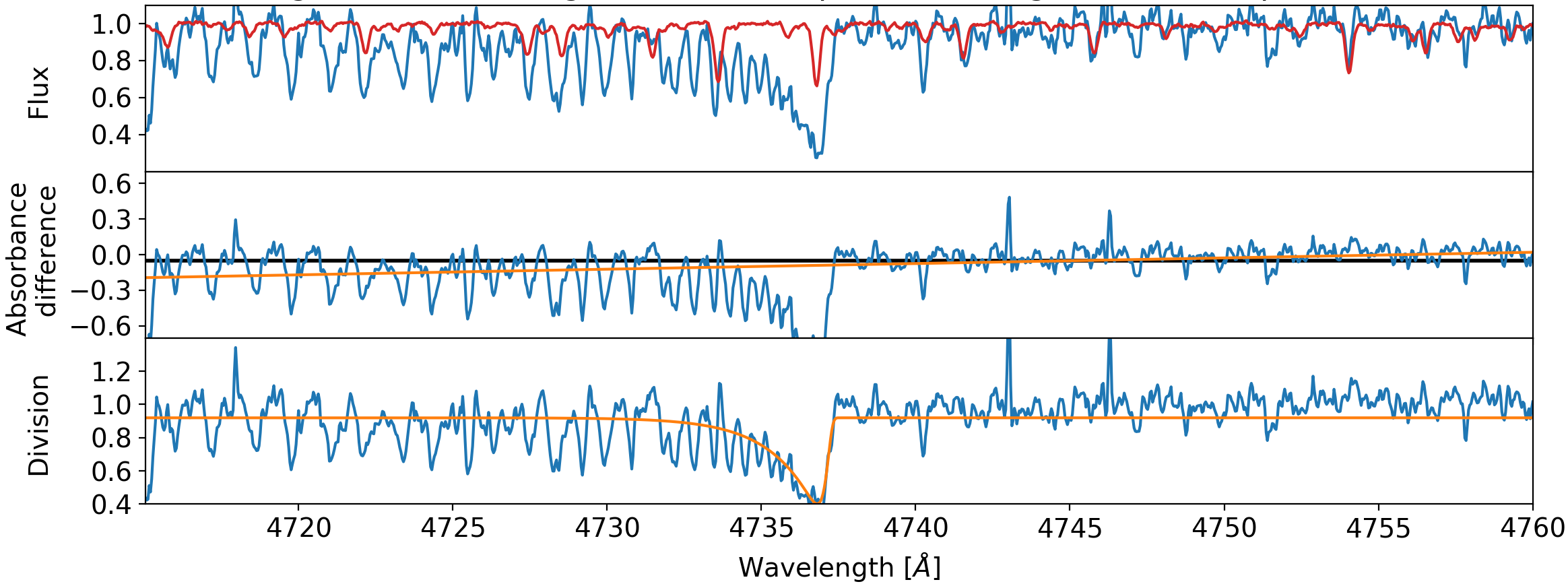}
	\caption{Example of a metal-poor carbon-enhanced candidate with strong Swan absorption feature at 4737~\AA, caused by the carbon C$_2$ molecules. The first panel shows the stellar spectrum in blue and a corresponding reference median spectrum in red. The reference median spectrum was computed as the per-pixel median flux value of spectra with similar stellar parameters as the spectrum shown in blue. The spectral difference (second panel) and division (third panel) were computed from those two spectra. The middle panel shows in orange a linear fit to the spectral difference that was used to identify spectra with the reduction problems on the blue border of the spectral range. The median value of the spectral difference is given by the black horizontal line. The orange curve in the last panel shows a fit that was used to determine the strength of the observed carbon feature. The shown spectrum belongs to a star with a Two Micron All-Sky Survey (2MASS) identifier J11494247+0051023 and iron abundance \Feh\ of $-1.17$, as determined by \TC.}
	\label{fig:carbon_example}
\end{figure*}

Carbon enhancement is observable in spectra as a strong additional absorption feature (Figure \ref{fig:carbon_example}) that is the strongest at the wavelength of the band's head. After that its power gradually decreases with decreasing wavelength. The most prominent and accessible for all of the spectra is a feature located at 4737~\AA, produced by a $^{12}$C$^{12}$C molecule. If other carbon features, like the one produced by a $^{13}$C$^{12}$C molecule at 4745~\AA\ (shown in Figure \ref{fig:carbon_example2}) are present in the spectrum, the carbon isotope ratio $^{12}$C/$^{13}$C in a star can be determined. Its determination was not attempted in the scope of this paper.

Detection of spectral features was tackled using two different classification procedures. First, a supervised procedure was used to identify the most prominent spectra with carbon enhancement. It is based on the assumption that we know where in the spectra those features are located and how they behave. This was augmented with an unsupervised dimensionality reduction algorithm that had no prior knowledge about the desired outcome. The goal of a dimensionality reduction was to transform n-dimensional spectra onto a 2D plane where differences between them are easier to analyse. The unsupervised algorithm was able to discern the majority of carbon-enhanced spectra from the rest of the data set and enabled us to discover spectra with less prominent carbon enhancement features.

\subsection{Supervised classification}
\label{sec:supervised}
To search for additional absorption features that are usually not found in spectra of chemically normal stars, we first built a spectral library of median spectra based on a rough estimates of stellar physical parameters derived by the automatic reduction pipeline, described in detail by \citet{2017MNRAS.464.1259K}. The median spectrum for every observed spectrum in our data set was computed from physically similar spectra with stellar parameters in the range of $\Delta$\Teff=$\pm75$~K, $\Delta$\Logg=$\pm0.125$~dex and $\Delta$\Feh=$\pm0.05$~dex around the stellar parameters of the investigated spectrum. The median spectrum was calculated only for observed targets with at least 5 similar spectra in the defined parameter range and with minimal SNR=15 per resolution element, as determined for the blue spectral arm. All considered spectra were resampled to a common wavelength grid with 0.04~\AA\ wide bins and then median combined. The normalisation of the spectra along with the radial velocity determination and the corresponding shift to the rest frame was performed by the automatic reduction pipeline \citep{2017MNRAS.464.1259K}. We checked that spectral normalisation and radial velocity determination are adequate also for carbon-enhanced stars. The normalisation procedure is done using a polynomial of low-order that is not strongly affected by the Swan band features or similar spectral structures. The radial velocity of a star is determined as an average of radial velocities that were independently determined for the blue, green, and red spectral arm. If one of the arms has a radial velocity deviating for more than two times the difference between the other two, it is excluded from the average \citep[further details in][]{2017MNRAS.464.1259K}. Therefore the final velocity should be correct even if one of those arms contains features that are not found in the set of reference spectra used in the cross-correlation procedure.

With the limitation of at least 5 spectra used for the computation of the median spectrum, some possibly carbon-enhanced stars, could be excluded from the supervised classification. The final number of spectra analysed by this method was 558,053.

Spectra, for which we were able to determine the median spectrum of physically similar objects, were analysed further. In the next step, we tried to determine possible carbon enhancement by calculating a flux difference and flux division between the observed stellar and median spectra, as shown in Figure \ref{fig:carbon_example}.

In order to describe the position, shape, and amplitude of the Swan feature with its head at 4737~\AA, we fitted a function that is based on a Log Gamma ($\log{}\Gamma$) distribution. The distribution, with three free parameters, was fitted to the division curve, where the Swan feature is most pronounced. Division curve, shown in the bottom panel of Figure \ref{fig:carbon_example}, was computed by dividing observed spectrum with its corresponding median spectrum. The fitted function $f$ can be written as:
\begin{equation}
\centering
f(\lambda) = f_0 - \log{}\Gamma(\lambda, S, \lambda_0, A).
\label{equ:loggamma}
\end{equation}
The shape of the curve is defined by an offset $f_0$, shape parameter $S$, centre wavelength $\lambda_0$, and amplitude $A$ of $\log{}\Gamma$ distribution, where $\lambda$ represents rest wavelengths of the observed spectrum. This function was selected because of its sharp rise followed by the gradual descent that matches well with the shape of a residual absorption observed in the Swan regions. The steepness of the rising part is determined by the parameter $S$ (lower value indicates steeper raise) and its vertical scaling by the parameter $A$. We are not aware of any other profile shapes used for fitting Swan bands in the literature.

To narrow down possible solutions for the best fitting curve, we used the following priors and limits. The initial value for the parameter $f_0$ was set to a median of all pixel values in the division curve and allowed to vary between $0.5$ and $1.5$. The limiting values are however newer reached. The centre of the $\log{}\Gamma$ distribution $\lambda_0$ was set to $4737$~\AA\ and was allowed to vary by $2$~\AA. Wavelength limits were set to minimise the number of mis-fitted solutions, where the best fit would describe the nearby spectral absorption lines not present in the median spectra or problematic spectral feature caused by the spectral data reduction as shown by Figure \ref{fig:bad_fit1}. We did not set any limits on parameters $A$ and $S$ in order to catch fitted solutions describing a spectrum difference that is different from the expected shape of the molecular absorption band.

\begin{figure*}
	\centering
	\includegraphics[width=\textwidth]{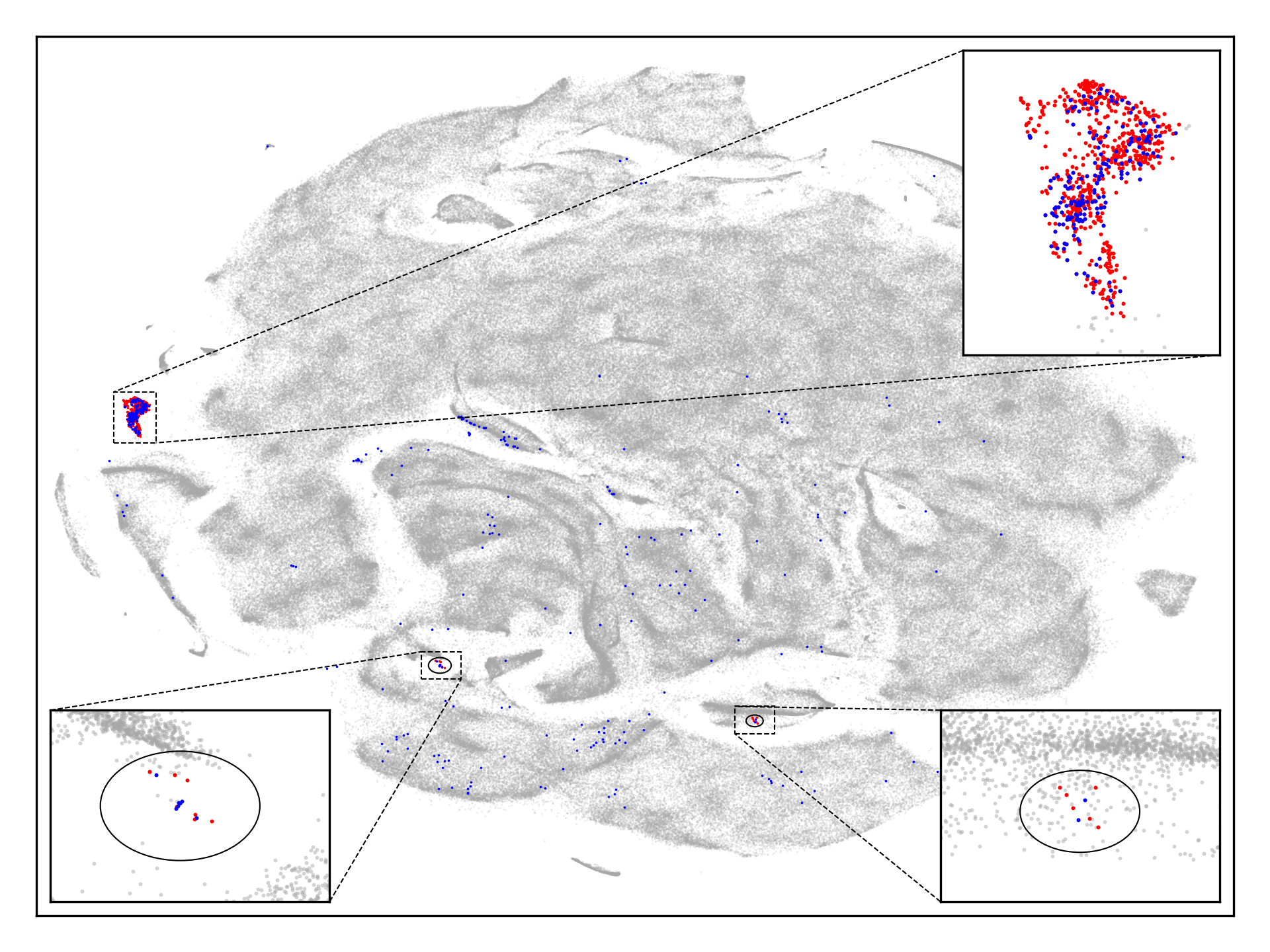}
	\caption{t-SNE projection of 588,681 observed spectra ranging between 4720 and 4890~\AA. Red dots (756 spectra) mark a clump in the projection that was manually selected to contain carbon-enhanced spectra. Superimposed blue dots represent carbon-enhanced spectra determined by the supervised algorithm. Outside the t-SNE selected clump, we have 224 spectra that were determined to be carbon-enhanced only by the supervised method. All other analysed spectra are shown in grey shades, depending on their density in the 2D projection. Two ellipses indicate regions where the majority of CEMP candidates is located in the projection.}
	\label{fig:tsne_plot}
\end{figure*}

By integrating the surface between the offset $f_0$ and the fitted curve we calculated the strength of the Swan band. The integral (\texttt{swan\_integ} in Table \ref{tab:out_table}) is derived between 4730 and 4738~\AA. It should not be used as a substitute for a carbon abundance measurement, but only to sort the detections of carbon-enhanced stars by their perceivable strength of the Swan band.

With so many spectra in our data set, unexpected reduction and analysis problems can hinder the selection of carbon-enhanced stars. In the first iteration, the results were ordered only by the value of the integrated Swan band region, but this proved to select too many spectra with reduction problems. Most of the problematic detections were caused by the incorrect normalisation of spectra with strong, non-carbon molecular bands. This is best observable at the border of a spectral range, where Swan bands are located in the case of HERMES spectra. There, normalisation can be poorly defined in the case of numerous nearby absorption lines. In order to prevent miss-detections, additional limits on the shape ($S <= 1$) and amplitude ($A <= 1$) of the $\log{}\Gamma$ distribution were used to filter out faulty fitting solutions. Figure \ref{fig:bad_fit2} represents one such example where the function $f(\lambda)$ was fitted to the absorption lines of a double-lined spectroscopic binary, producing a shape of the function that is not characteristic for the analysed molecular band head. To remove spectra with reduction problems or peculiarity that would result in wrongly determined strength of the Swan band, we are also analysing the slope of the spectral difference and its integral in the limits of the Swan bands. One of the spectral trends that we are trying to catch with those indicators is shown in Figure \ref{fig:bad_fit2}, where spectral difference and its linear fit are steeply rising at the border of the spectrum.

By visual inspection of the algorithm diagnostic plots shown in Figure \ref{fig:carbon_example}, we limited a final selection to 400 spectra with the strongest carbon enhancement that was still visually recognisable. The last selected spectrum is shown in the Figure \ref{fig:carbon_last_supervised}. Selection of spectra with lower enhancement, would introduce possibly wrong classification of stars whose enhancement is driven by spectral noise levels, data reduction or any other process that has subtle effect on the spectral shape.

\subsection{Unsupervised classification}
\label{sec:unsupervised}
With numerous spectra of different stellar types, chemical composition, and degree of carbon enhancement, some of them might show different carbon features or be insufficiently distinctive to be picked out by the above supervised algorithm.

Another analysis technique, which is becoming increasingly popular is a dimensionality reduction procedure named t-distributed Stochastic Neighbor Embedding \citep[t-SNE,][]{van2008visualizing}, that has already proved to be beneficial in comparison and sorting of unknown spectral features of the same data set \citep{2017ApJS..228...24T}. This is done by projecting the complete spectra onto a 2D plane by computation of similarities between all pairs of investigated spectra. It has been shown that the algorithm can cluster and distinguish spectra with absorption or emission features. The algorithm arranges spectra in a 2D plane, such that it clusters similar spectra together based on their similarity measure. As the transformation is variable and non-linear, the actual distance between two objects in a final 2D plane does not linearly depend on the spectral similarity measure. This property of the t-SNE algorithm ensures more homogeneous coverage of the 2D plane in comparison to other dimensionality reduction methods.

The t-SNE projection shown in Figure \ref{fig:tsne_plot} was computed from normalised spectra between 4720 and 4890~\AA. To maximise the number of analysed spectra, no other limiting cuts than the validity of the wavelength solution \citep[bit 1 in \texttt{red\_flag} set to 0 by reduction pipeline, ][]{2017MNRAS.464.1259K} in this arm was used. This resulted in 588,681 individual spectra being analysed by the automatic unsupervised algorithm. This is $\sim30$k more spectra than in the case of supervised classification, where we applied more strict criteria for the selection of analysed spectra (Section \ref{sec:supervised}). 

Without any prior knowledge about the location of objects of interest in the obtained projection, we would have to visually analyse every significant clump of stars in order to discover whether the carbon-enhanced population is located in one of them. This can be simplified by adding the results of the supervised classification into this new projection. In Figure \ref{fig:tsne_plot}, the stars identified by the supervised classification are shown as blue dots plotted over grey dots representing all spectra that went into the analysis. The majority of blue dots are located in a clump on the left side of the projection. A high concentration of objects detected by a supervised method leads us to believe, that this isolated clump represents carbon-enhanced objects in the t-SNE projection. To select stars inside the clump, we manually drewn a polygon around it.

Inspection of other blue labelled spectra outside the main clump revealed that their slight carbon enhancement could not be identified by the t-SNE similarity metric as the spectra comparison might have been dominated by another spectral feature.

Additional exploration of the t-SNE projection revealed two smaller groups of metal-poor carbon-enhanced spectra located inside ellipses shown in Figure \ref{fig:tsne_plot}. A confirmation that those regions are populated with metal-poor stars can be found in Figure \ref{fig:tsne_teff_feh} where the dots representing spectra in the projection are colour coded by \Feh\ and \Teff. To maximise the number of those objects in the published catalogue, we manually checked all undetected spectra in the vicinity of the detected ones. This produced additional 13 CEMP detections.

\subsubsection{t-SNE limitation}
\begin{figure}
	\centering
	\includegraphics[width=\columnwidth]{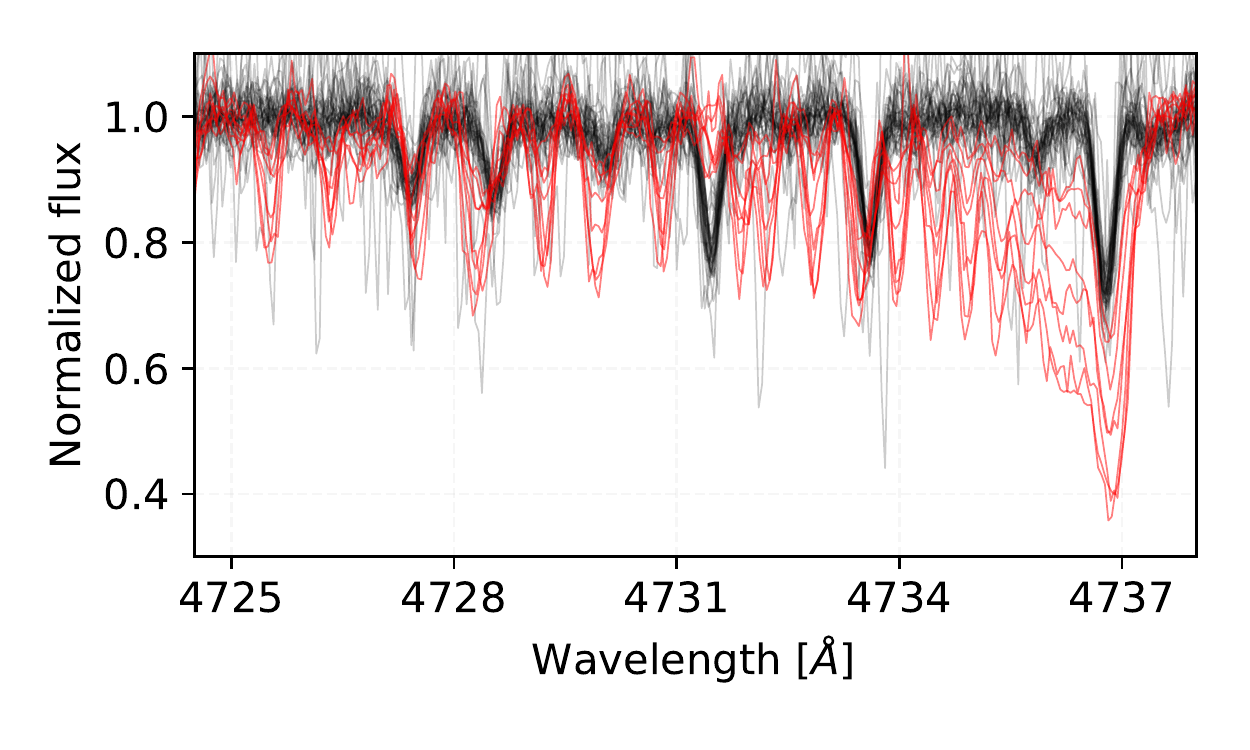}
	\caption{A collection of spectra that were determined to be mutually very similar by the t-SNE algorithm. Out of 46 spectra inside the right black ellipse in Figure \ref{fig:tsne_plot} we identified 8 carbon-enhanced spectra with visually very different and distinctive spectrum in the region from 4734 to 4737~\AA\ that is also depicted in this figure. For easier visual recognizability, they are coloured in red.}
	\label{fig:tsne_cemps}
\end{figure}
While checking the local neighbourhood of some of the blue dots in Figure \ref{fig:tsne_plot} that are strewn across the t-SNE projection we identified a possible limitation of our approach for the automatic detection of specific peculiar spectra if their number is very small compared to the complete data set. Figure \ref{fig:tsne_cemps} shows a collection of a few carbon-enhanced spectra embedded between other normal spectra that were taken out of the right ellipsoidal region in Figure \ref{fig:tsne_plot}. As they are quite different from the others they were pushed against the edge of a larger cluster in the projection, but their number is not sufficient to form a distinctive group of points in the projection. Therefore any automatic algorithm that would try to distinguish those objects based solely on a local density of points would most probably fail.

\begin{figure}
	\centering
	\includegraphics[width=\columnwidth]{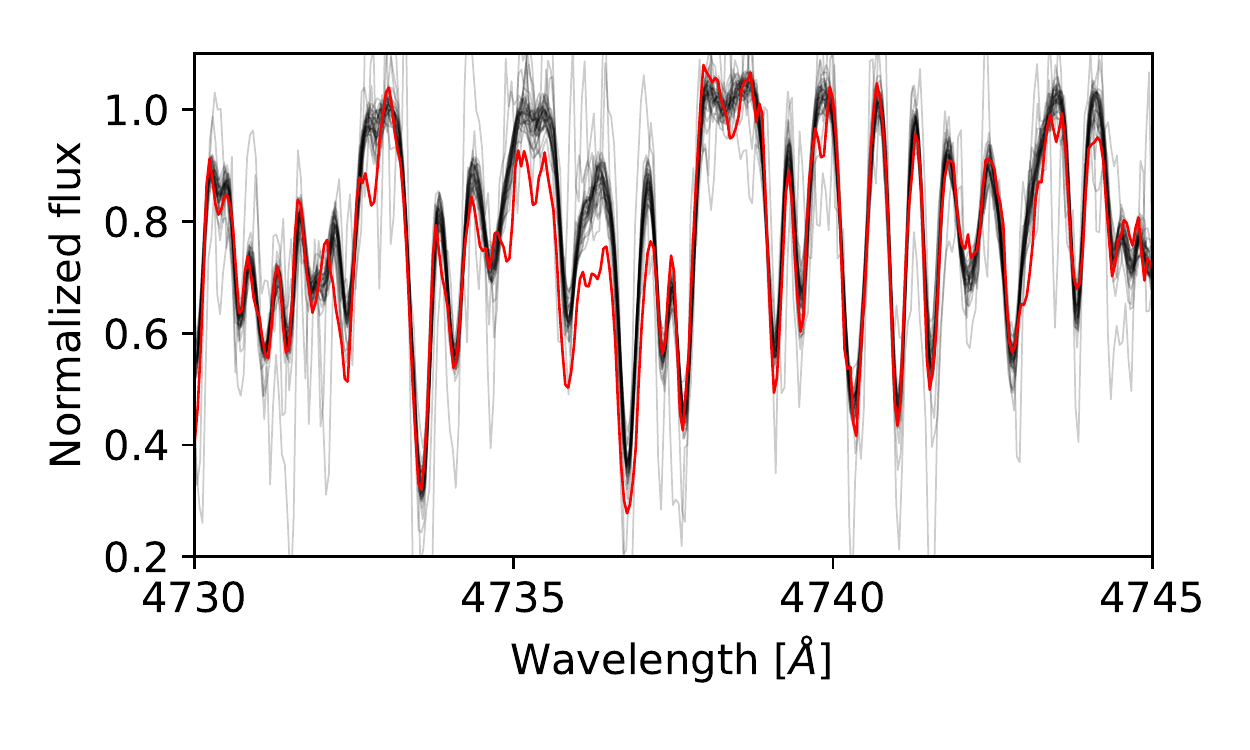}
	\caption{Spectral comparison between one of the detected carbon-enhanced stars in red and its 30 closest neighbours in the t-SNE projection shown as black curves. Enhancement in the spectrum was probably not sufficiently distinct and was dominated by the spectral noise. Therefore the spectrum was placed among other physically similar spectra without visible enhancement.}
	\label{fig:tsne_30close}
\end{figure}

Another specific of the t-SNE projection that we must be aware of is how it computes the similarity between analysed spectra. Combined similarity, which is computed as a sum of per pixel differences, has zero knowledge about the location where in the spectrum those differences occur. The red spectrum in Figure \ref{fig:tsne_30close} with a slight signature of carbon enhancement in the range between 4734 and 4737~\AA\ has been placed among spectra with similar physical properties. Its slight carbon enhancement and comparable spectral noise to other spectra in its vicinity are probably the reason why it was placed in such a region of the t-SNE projection. This could be solved by using a smaller portion of the spectrum in a dimensionality reduction, which could at the same time lead to a loose of other vital information about a star.

\section{Candidate characteristics}
\label{sec:analysis}
The final list of detected carbon-enhanced stars consists of 918 stars, corresponding to 993 spectra detected by at least one of the described methods. Among them, 63 stars were observed and identified at least twice and up to a maximum of four times. Those identifications belong to repeated observations that were performed at different epochs. Because not all of the observed spectra were considered in the classification procedure (due to the limitations described in Section \ref{sec:classification}) this is not the final number of stars with repeated observations. By searching among the complete observational data set, the number of carbon-enhanced stars with repeated observations increases to 90.

Out of those 90 stars, every repeated observation of 56 stars was classified as being carbon-enhanced. In total, we detected $76.5$~\% of the carbon-enhanced spectra among repeated observations where at least one of the repeats have been classified as having enhanced carbon features in its spectrum. The unclassified instances usually have a low SNR value that could decrease their similarity value towards other carbon-enhanced stars in the t-SNE analysis or have incorrect stellar parameters and were therefore compared to an incorrect median spectra during the supervised analysis.

\subsection{Radial velocity variations}
\label{sec:binaries}
\begin{figure}
	\centering
	\includegraphics[width=\columnwidth]{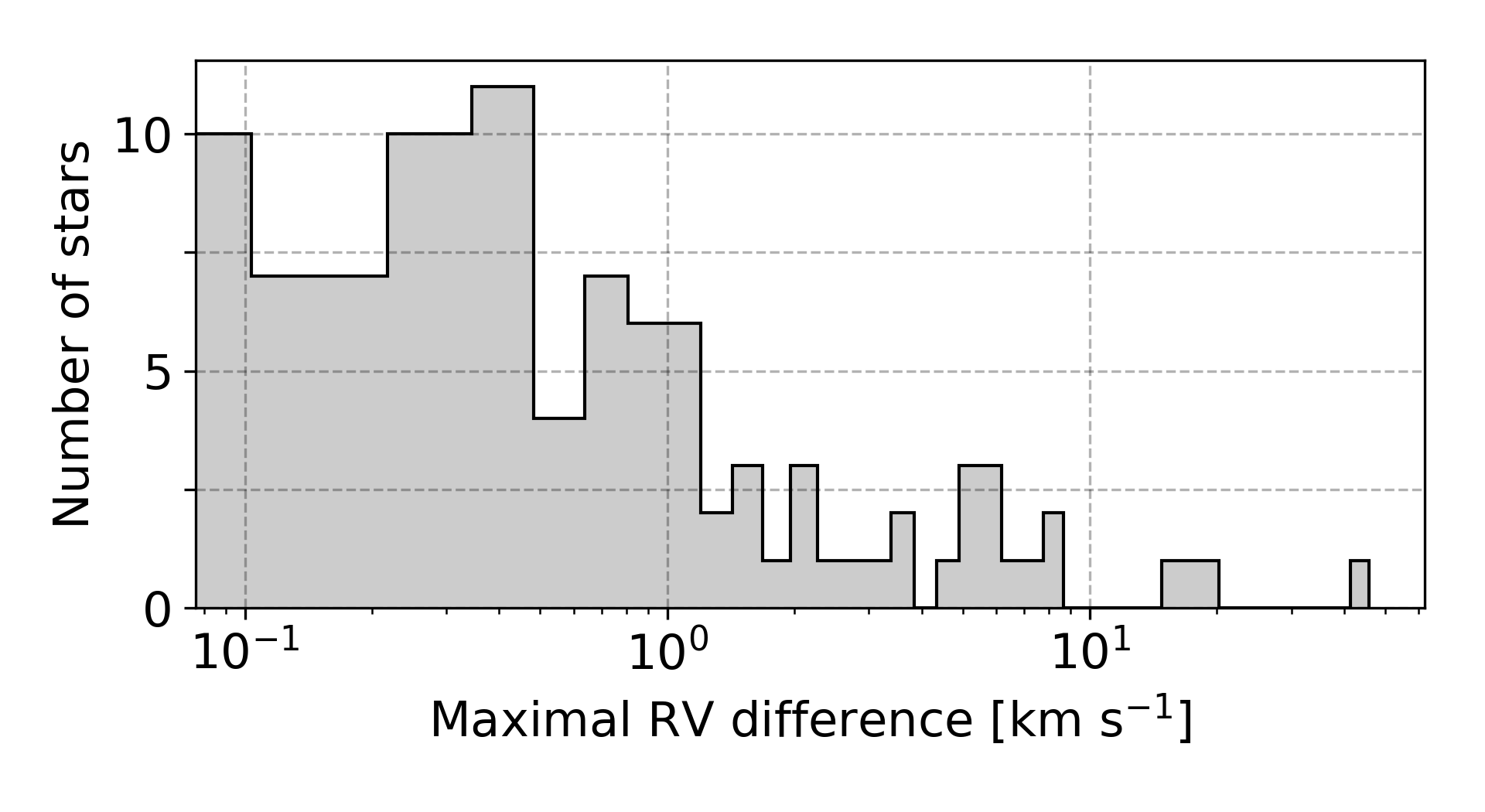}
	\caption{Distribution of maximal velocity change between repeated observations of the stars that were classified as carbon-enhanced.}
	\label{fig:rv_rep_dist}
\end{figure}

With repeated observations in the complete observational data set, we can look into measured radial velocities and investigate a number of possible variables that should be high for certain types of carbon-enhanced objects \citep{2016ApJ...826...85S}. Taking into account all of the repeated observations in our data set and not just the repeats among the identified spectra, 52 out of 90 stars show a minimum velocity change of $0.5$~\kms (70 stars with minimum change of $0.25$ \kms) and a maximum of $45$~\kms in different time spans ranging from days to years. The detailed distribution is presented by Figure \ref{fig:rv_rep_dist}. That kind of change can hint at the presence of a secondary member or at intrinsic stellar pulsation \citep{2002AA...390..987B, 2010JApA...31..177L, 2012A&A...544A..10B}, as carbon-enhanced stars are found among all long period variable classes \citep[Mira, SRa, and SRb, ][]{2013A&A...553A..93B, 2014A&A...568A.100B}. Follow-up observations are needed to determine their carbon sub-class and subsequently the reason behind variations of radial velocity.

Visual inspection of variable candidates revealed that none of them shows obvious multiplications of absorption spectral lines, a characteristic of a double-lined binary system. Therefore we can conclude that none of them is a binary member in which both components are of comparable luminosity and a difference between their projected radial velocities is high enough to form a double-lined spectrum. From our simulations with median spectra, such line splitting becomes visually evident at the velocity difference of $\sim$14~\kms. If the components do not contribute the same amount of flux, the minimal difference increases to $\sim$20~\kms.

Chemical peculiarity of a dwarf carbon-enhanced star (dC) that exhibits enhancement of C$_2$ in its spectra could be explained by its interaction with a primary star in a binary system \citep{2018ApJ...856L...2M}. Chemically enhanced material is thought to be accreated from the evolved AGB companion. Less than thirty of such systems, that show signs of the existence of an invisible evolved companion who might have enriched a dC by the carbon, have been identified spectroscopically to date \citep{1986ApJ...300..314D, 2018ApJ...856L...2M, 2018MNRAS.479.3873W}, giving us the possibility to greatly increase the list with every additional confirmed object. The only detected dC star (for criteria see Section \ref{sec:cannon_params}) with repeated observations shows that its radial velocity is unchanged on the order of $0.1$~\kms\ during the 2 years between consecutive observations. Hence, it cannot be classified as a possible binary system from those two observations alone. The lack of a clear evidence for binarity among dC stars, especially among the most metal-poor, can also be explained by another enrichment mechanism. \citet{2018MNRAS.477.3801F} showed that a substantial fraction of those stars belongs to the halo population based on their kinematics information. Combined with the results of \citet{2016ApJ...833...20Y} that classified the prototype dC star \mbox{G 77-61} as a CEMP-no star, that are known to have intrinsically low binarity fraction \citep{2014MNRAS.441.1217S, 2016A&A...586A.160H}, their carbon-enhancement may be of a primordial origin.

\begin{figure}
	\centering
	\includegraphics[width=\columnwidth]{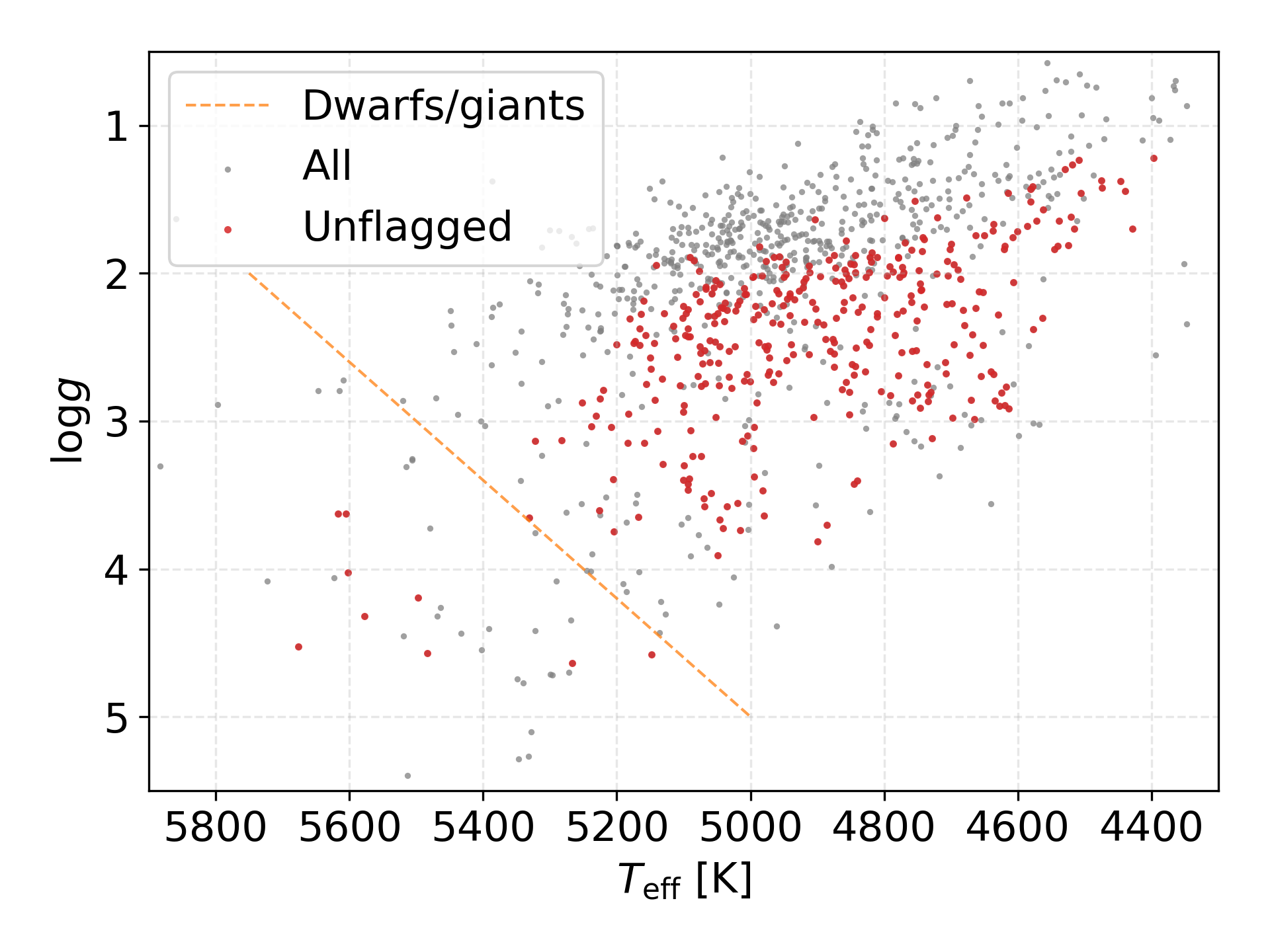}
	\caption{Kiel diagram for a subset of 338 detected carbon-enhanced stars with valid stellar parameters in red. Uncertain positions of flagged stars are shown with grey dots. Dashed orange line illustrates the border between giants and dwarfs.}
	\label{fig:kiel_plot}
\end{figure}

\subsection{Stellar parameters}
\label{sec:cannon_params}
For the analysis of stellar parameters, we used values determined by \TC\ data interpolation method that was trained on actual observed HERMES spectra. To exclude any potentially erroneous parameter, we applied a strict flagging rule of \texttt{flag\_cannon}=0 \citep[an extensive description of flagging procedure can be found in][]{buder2018}, thus obtaining a set of 347 objects with trustworthy stellar parameters. Such a large percentage of flagged objects could be attributed to their nature as an additional elemental enhancement that we are looking for might not be a part of the training set. A raised quality flag would hint that the spectrum is different from any other in the training set or that the fit is uncertain and has a large $\chi^2$. Therefore flagged parameters have to be used with care, especially on the border of, and outside the training set.

The majority (338) of the unflagged detected objects are giants and only 9 are confirmed to be dwarf stars based on their spectroscopic stellar parameters (Figure \ref{fig:kiel_plot}).

\begin{figure*}
	\centering
	\includegraphics[width=\textwidth]{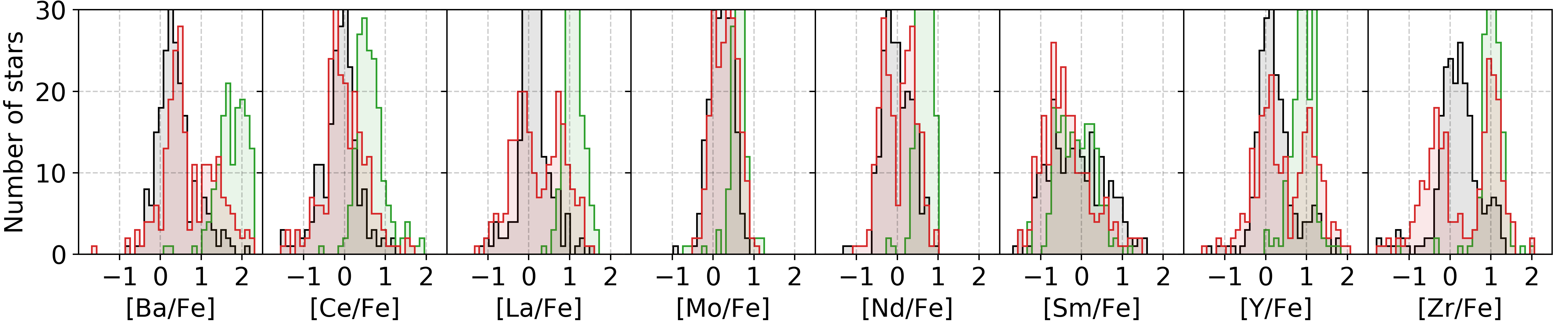}
	\caption{Distribution of s-process element abundances for stars in three different groups. The most enhanced group in green represent carbon-enhanced stars located in the t-SNE selected clump of stars. The red distribution presents all other detections that are placed around the projection, and outside the clump. As a control group, the same distribution in black is shown for their closest t-SNE neighbours, therefore the black and red distribution contain an equal number of objects. No abundance quality flags were used to limit abundance measurements.}
	\label{fig:sprocess_hist}
\end{figure*}

\subsection{S-process elements}
\label{sec:sprocess}
Focusing on a spectral signature of the detected objects inside and outside the t-SNE selected clump (Figure \ref{fig:tsne_plot}) we can further investigate which spectral feature might have contributed to their separation. The distributions of their abundances in Figure \ref{fig:sprocess_hist} and strength of spectral features corresponding to the same elements in Figure \ref{fig:sprocess_spectrum} hints to an enhancement of s-process elements among stars inside the selected clump. This additional enhancement might be another reason, besides the carbon enhancement, for the algorithm to cluster all of those stars as being different from the majority of spectra. 

\begin{figure}
	\centering
	\includegraphics[width=\columnwidth]{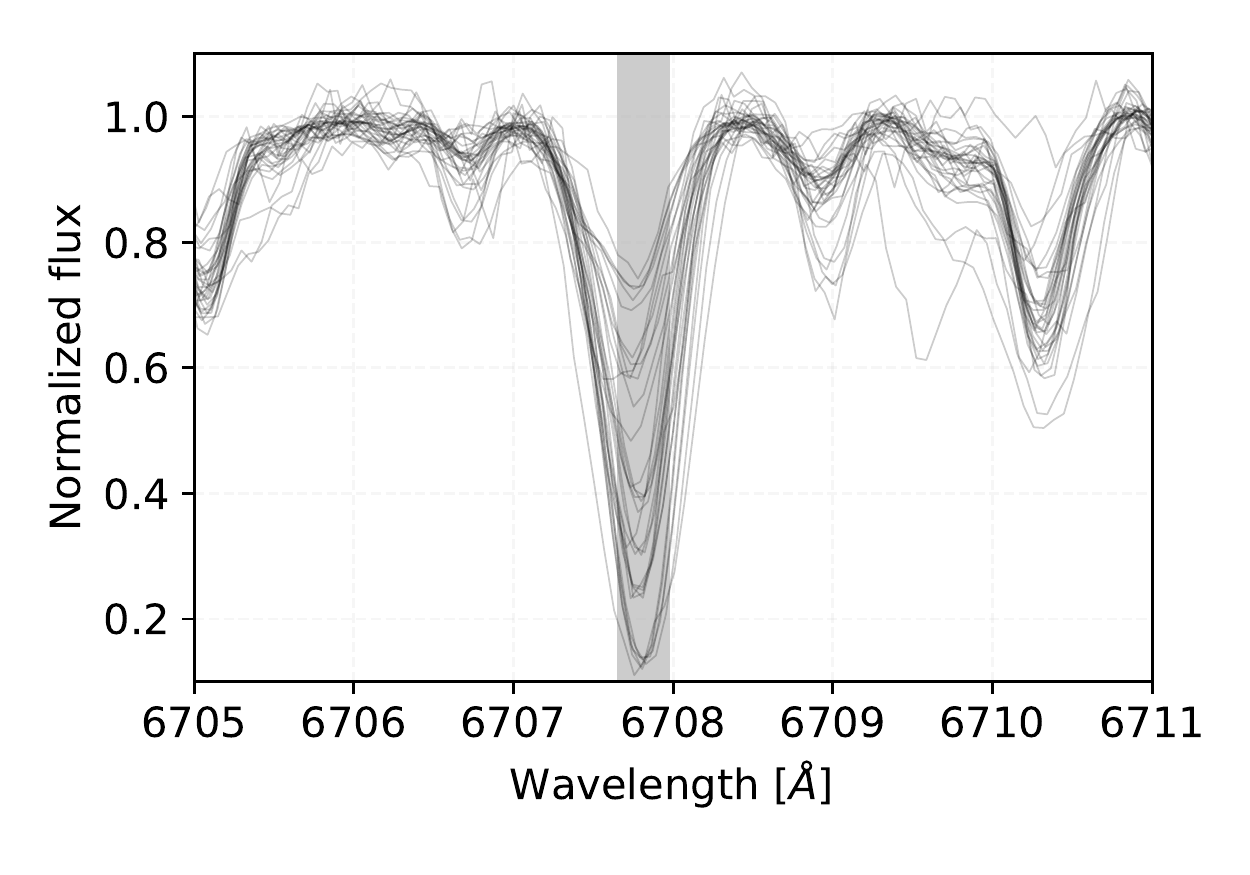}
	\caption{Spectral subset of 32 lithium-rich carbon-enhanced stars among the identified stars. The highlighted wavelength region is used by \TC\ to determine the lithium abundance of a star.}
	\label{fig:li_abund}
\end{figure}	

\subsection{Lithium abundance}
\label{sec:lithium}
The derivation of elemental abundances for known carbon-enhanced stars has shown that some of them can exhibit strongly enhanced levels of Li in their atmosphere \citep{1991A&A...245L...1A}. Lithium is thought to be produced by hot-bottom burning \citep{1974ApJ...187..555S} and brought to the surface from the stellar interior. Investigation of the Li line at 6707~\AA\ revealed 32 of such stars. Their spectra, centred around the Li feature, show a greatly varying degree of absorption in Figure \ref{fig:li_abund}.

\subsection{Sub-classes}
\label{sec:subclasses}
Following a revision of the original MK classification \citep{1941ApJ....94..501K} introduced by \citet{1996ApJS..105..419B}, carbon stars are separated into five different classes named \mbox{C-H}, \mbox{C-R}, \mbox{C-J}, \mbox{C-N}, and Barium stars. Of all the spectral indices proposed for the spectral classification, we are only able to measure a small part of Swan C$_2$ bands and Ba II line at 6496~\AA. For a more detailed classification of detected objects into proposed classes, we would need to carry out additional observations with a different spectroscopic setup to cover all the significant features. 

Additionally, the features caused by the $^{13}$C$^{12}$C molecule are strongly enhanced only for a handful of spectra in our data set, therefore we did not perform any isotopic ratio analysis or identification of possible C-J objects, which are characterised by strong Swan bands produced by the heavier isotopes.

According to the abundance trends presented in Section \ref{sec:sprocess} and the classification criteria defined by \citet{1996ApJS..105..419B}, we could argue that the stars selected from the t-SNE projection belong to the C-N sub-class. Their s-process elements are clearly enhanced over Solar values (Figure \ref{fig:sprocess_hist}), but the actual values should be treated with care as they are mostly flagged by \TC. This uncertainty might come from the fact that the training set does not cover carbon-enhanced stars and/or stars with such enhancement of s-process elements.

\subsection{Match with other catalogues}
In the literature we can find numerous published catalogues of carbon-enhanced (CH) stars \citep{2001A&A...375..366C, 2001BaltA..10....1A,2016ApJS..226....1J} and CEMP stars \citep{2007ApJ...658..367K, 2010A&A...509A..93M, 2010AJ....139.1051P, 2014ApJ...797...21P, 2015A&A...581A..22A, 2017yCat..18330020Y} observed by different telescopes and analysed in inhomogeneous ways. Most of those analyses were also performed on spectra of lower resolving power than the HERMES, therefore some visual differences are expected for wide molecular bands. By matching published catalogues with the GALAH observations that were analysed by our procedures, we identified 44 stars that matched with at least one of the catalogues. Of these, 28 were found in CH catalogues and 16 in CEMP catalogues.

From the stars recognised as CEMPs in the literature, we were able to recover only 1 of them. Visual assessment of the diagnostic plots provided by our analysis pipeline proved that the remaining 15 CEMP matches do not expresses any observable carbon enhancement in Swan bands and were therefore impossible to detect with the combination of our algorithms. The reason for this difference between our and literature results might be in the CEMP selection procedure employed by the aforementioned literature. Every considered study selects their set of interesting stars from one or multiple literature sources based on values of \Meh\ and \Cfe\ that were measured from the atomic spectral lines and not molecular lines. 

\begin{figure}
	\centering
	\includegraphics[width=\columnwidth]{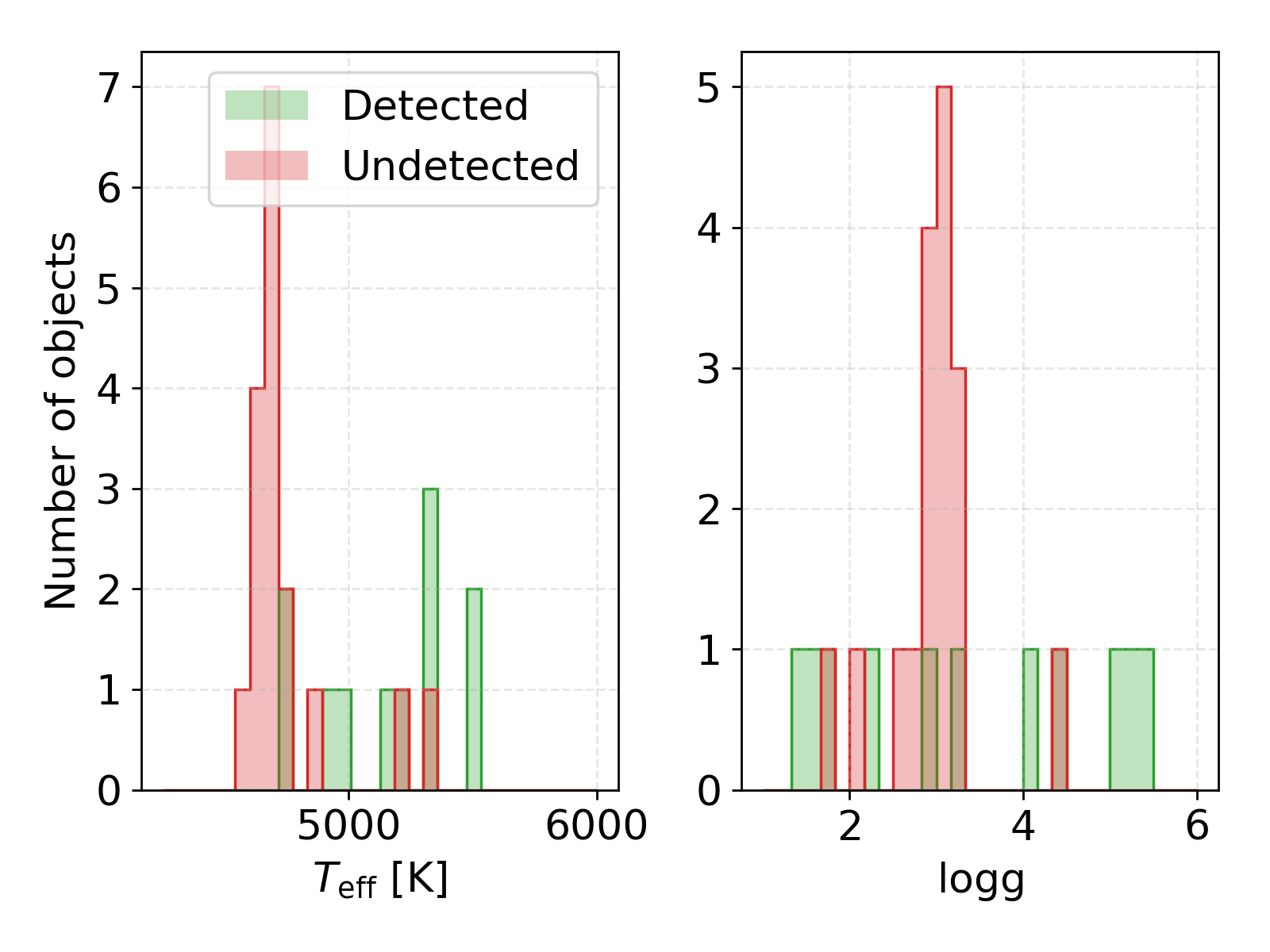}
	\caption{Comparison between the stellar parameters of detected (green histogram) and undetected (red histogram) carbon-enhanced stars found in literature.}
	\label{fig:ch_xmatch}
\end{figure}

The match is larger in the case of CH matches, where we were able to confirm 11 out of 33 possible matched carbon-enhanced stars. As the observed molecular bands are prominent features in the spectra, we explored possible reasons for our low detection rate. Visual inspection of spectra for the remaining undetected matched stars proved that they also show no or barely noticeable carbon enhancement in the spectral region of Swan bands, therefore reason must lie in the detection procedures used in the cited literature. \citet{2001A&A...375..366C} used low-resolution spectra to evaluate enhancement of C$_2$ and CN bands. The results are also summarised in their electronic table. In here, all of our undetected stars are marked to contain enhanced CN bands but no C$_2$ bands. Combining this with Figure \ref{fig:ch_xmatch} we speculate that those stars occupy a narrow range of parameter space where C$_2$ is not expressed and therefore undetectable in the HERMES spectra. 

Number of successfully detected stars matched between the surveys could also be influenced by different excitation temperatures of analysed carbon-rich molecules. Frequently studied photometric G-band, that is not present in our spectra, covers a spectral region rich in CH molecule features whose temperature dependence is different than for a C$_2$ molecule. Presence of those bands is identified by classifying a carbon-enhanced star into C-H sub-class (see Section \ref{sec:subclasses}). As we detected all C-H stars identified by \citet{2016ApJS..226....1J}, that are also present in the GALAH data set, we are unable to discuss about the selection effect in the \Teff\ range between $\sim$5100 and $\sim$5300~K where those three stars were found.

The position of all stars matched with the literature is also visualised on the \mbox{t-SNE} projection in Figure \ref{fig:tsne_ref_ch}, where it can be clearly seen that they lie outside the selected clump with identified carbon enhancement and are strewn across the projection. Close inspection of spectra that are spatially near the aggregation of CEMP stars from the literature, revealed no visible carbon enhancement. The enhancement is present neither in form of molecular bands nor expressed as stronger atomic carbon line. They therefore are indistinguishable from other metal-poor stars with similar physical parameters.

\begin{figure}
	\centering
	\includegraphics[width=\columnwidth]{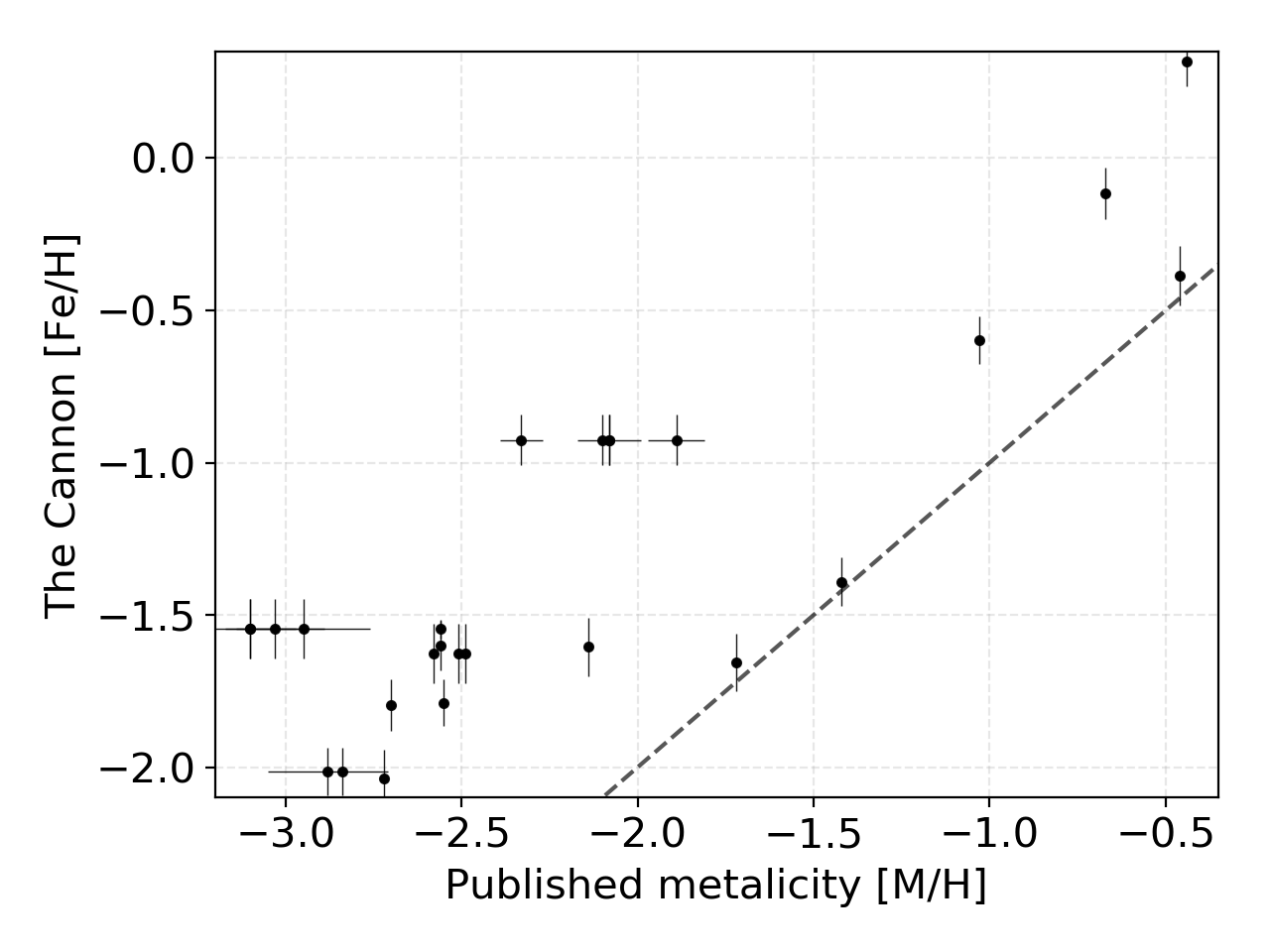}
	\caption{Correlation between published metallicities and \TC\ iron abundance for the stars that were classified as CEMPs in the literature. As some of those stars were taken from multiple literature sources, we also have multiple determinations of \Meh\ for them. This can be identified as horizontal clusters of dots at different \Meh, but with the same \Feh. Where available, uncertainties of parameters are shown. The dashed line follows a 1:1 relation.}
	\label{fig:cemps_feh}
\end{figure}

\section{Metal-poor candidates}
\label{sec:cemp}
CEMP stars are defined in the literature as having low metallicity \Meh \textless $-1$ and strong carbon enrichment \Cfe \textgreater $+1$. In the scope of this analysis, we assume that our measurement of \Feh\ is a good approximation for the metallicity. To be sure about this we compared \Meh\ values of CEMP stars found in the literature and \Feh\ derived by \TC\ for the same stars. The relation between them is shown in Figure \ref{fig:cemps_feh}. We see that our values start deviating from the published values at metallicities bellow $-1.5$. Bellow that threshold the differences are in the range of $\sim1$ dex, but the same trend is obvious for both data sets. The uncertainty of the published \Meh, derived from multiple sources, can reach up to $0.5$. 

\begin{figure}
	\centering
	\includegraphics[width=\columnwidth]{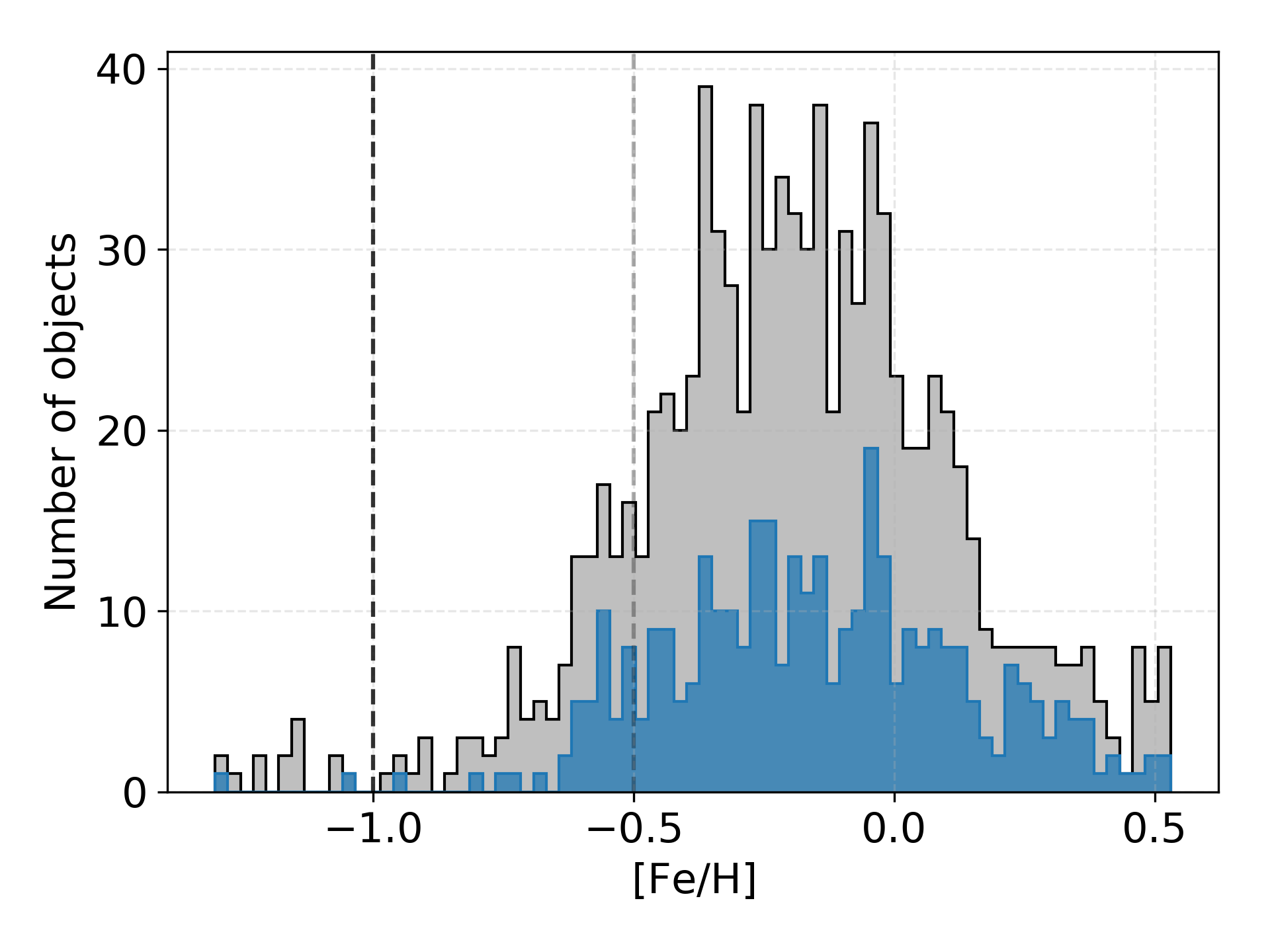}
	\caption{Histogram of \Feh\ for detected carbon-enhanced stars with valid \TC\ stellar parameters in blue and for every detected carbon-enhanced star in grey. Two vertical lines are located at iron abundances of $-1.0$ and $-0.5$.}
	\label{fig:feh_candidates}
\end{figure}

Taking unflagged \TC\ parameters and abundances of the detected objects we can determine possible CEMP candidates among our sample. As also shown by Figure \ref{fig:feh_candidates} our set of carbon-enhanced stars consists of 41 objects with \Feh \textless $-0.5$ and 2 objects with \Feh \textless $-1.0$. If we also include potentially incorrect parameters, the number of objects with \Feh \textless $-1.0$ increases to 28, which is equal to $2.8$~\% of detected carbon-enhanced spectra. In any case, none of them has a valid determination of carbon abundance. Analysing HERMES spectra in order to determine carbon abundance is difficult because the automatic analysis is based on only one very weak atomic absorption line that is believed to be free of any blended lines. Consequently, we are also not able to measure the \CO\ abundance ratio, as a majority of determined \Cfe\ abundances is flagged as unreliable. Complementary observations are needed to determine the abundance and confirm suggested CEMP candidates.

\begin{figure}
	\centering
	\includegraphics[width=\columnwidth]{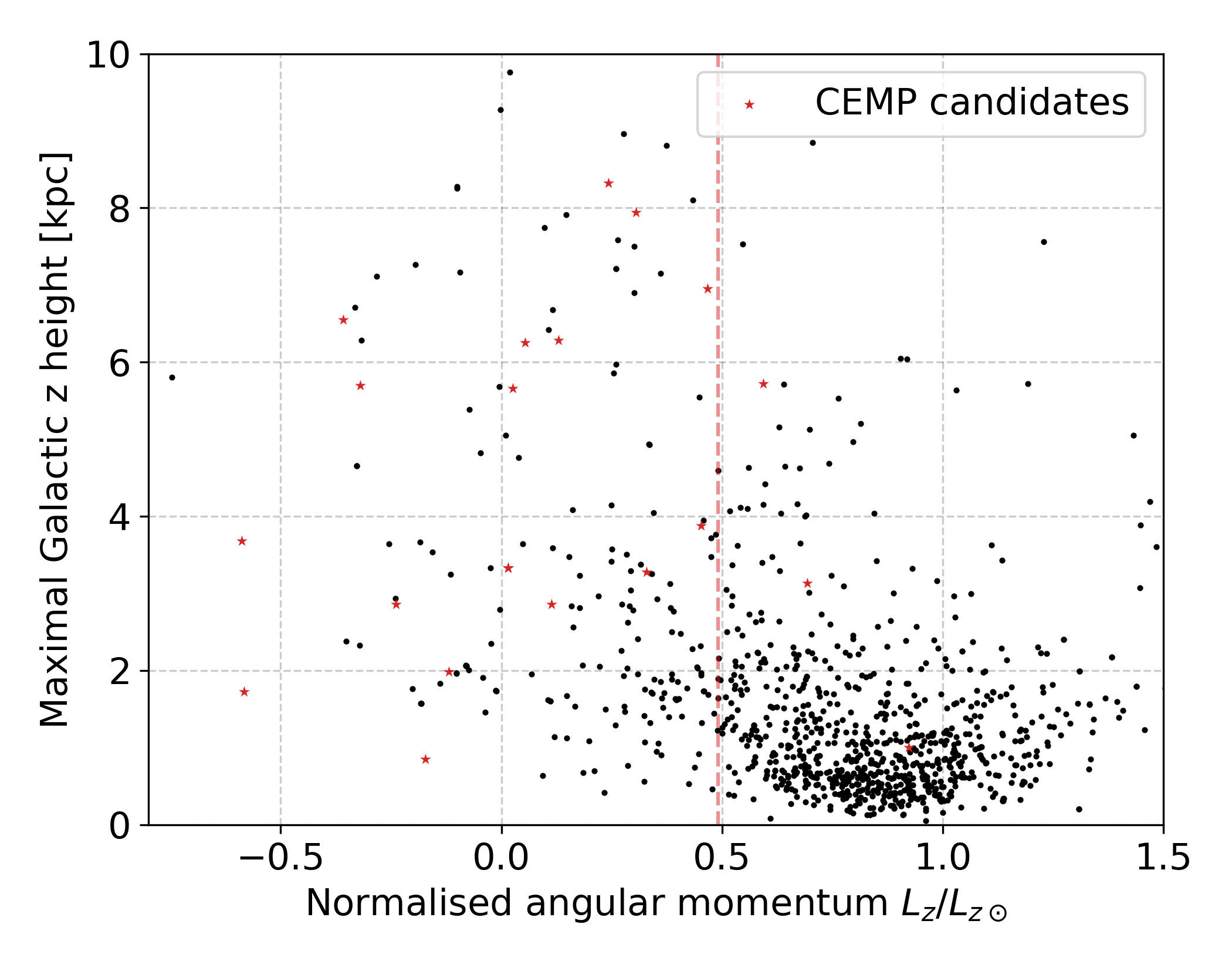}
	\caption{Distributions of maximal height above/below the Galactic plane reached by detected stars on their orbit around the centre of the Galaxy in comparison to their normalised angular momentum $L_z$. Vertical dashed line at $1000$~\kms~kpc highlights the transition from the halo to the disk population, where a majority of the halo stars is located below this threshold \citep[the threshold was visually estimated from similar plots in][]{2018ApJ...860L..11K}.  CEMP candidates are marked with star symbols.}
	\label{fig:orbits_zmax}
\end{figure}

A low number of metal-poor candidates could also be explained by the specification of the HERMES spectrograph as its spectral bands were not selected in a way to search for and confirm most metal-poor stars. With the release of {\it Gaia} DR2 data \citep{2018A&A...616A...1G}, stars low/high-metallicity could also be compared with their Galactic orbits. To determine the distribution of detected stars among different Galactic components, we performed an orbital integration in \texttt{MWPotential2014} Galactic potential using the \texttt{galpy} package \citep{2015ApJS..216...29B}. In order to construct a complete 6D kinematics information, {\it Gaia} parallax and proper motion measurements were supplemented with the GALAH radial velocities. Results shown in Figure \ref{fig:orbits_zmax} suggest that our CEMP candidates could belong to two different components of the Galaxy. Stars with maximal $z$~<~4~kpc most probably belong to the thick disk and stars with $z$~>~5~kpc to the halo population that is inherently metal-poor. This is also supported by their angular momentum in the same plot and their Galactic velocities shown in Figure \ref{fig:orbits_vxvyvz}.

When looking at the distribution of \Feh\ for the complete set of observed stars, we find a comparable distribution as for carbon-enhanced stars. Similarly, about $1.8$~\% of stars are found to be metal-poor with \Feh \textless $-1.0$.

\section{Follow-up observation}
\label{sec:asiago}

\begin{figure}
	\centering
	\includegraphics[width=\columnwidth]{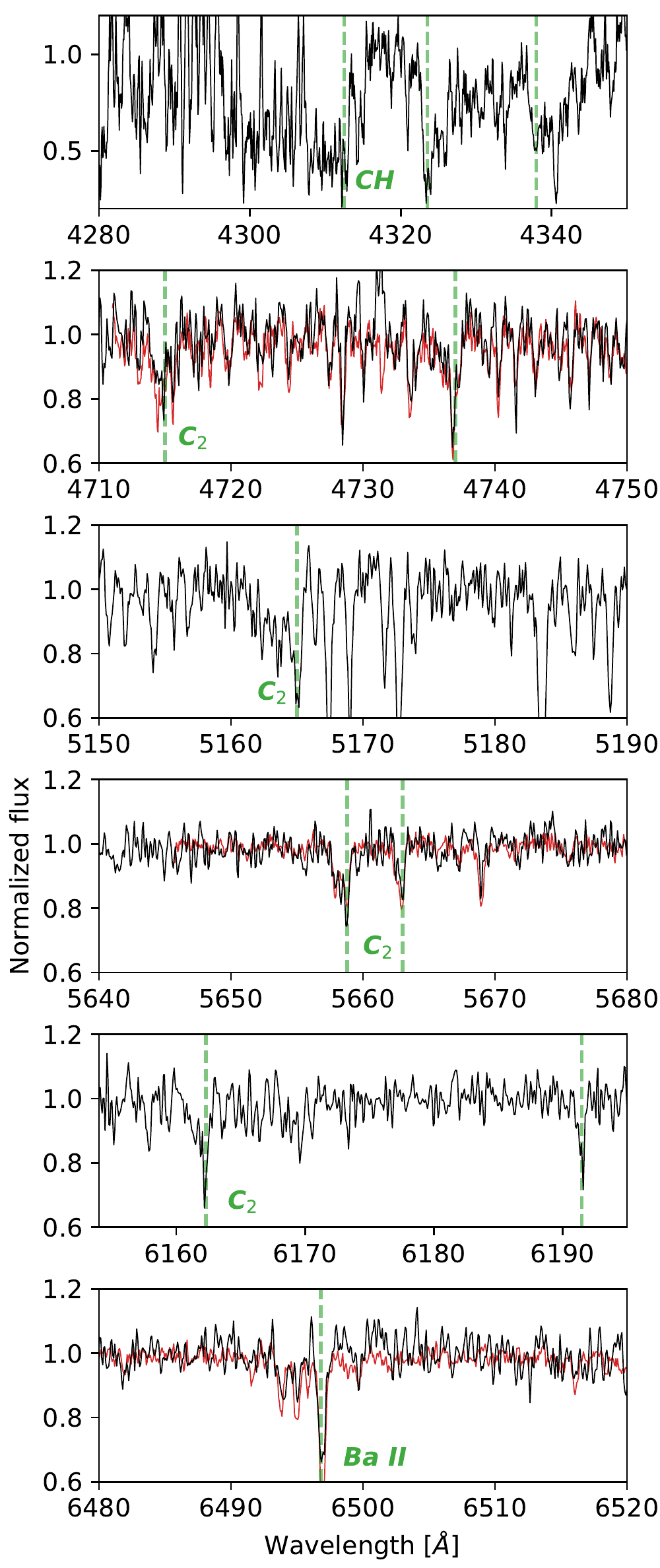}
	\caption{Subsets from the follow-up Asiago spectrum with resolving power comparable, but not identical, to the HERMES spectrum. It contains multiple spectral features used to evaluate carbon enhancement in a star and its carbon sub-class. Relevant spectral features are marked with vertical dashed green lines and labels that represent a molecule or element that is responsible for the features shown in the individual panel. The 2MASS identifier of the observed star is J11333341-0043060. In the wavelength ranges where it is available, the GALAH spectrum of the same star is shown in red.}
	\label{fig:asiago}
\end{figure}

To further classify and analyse one of the detected objects, a star with 2MASS identifier J11333341-0043060 was selected for a follow-up observation. We acquired its high-resolution Echelle spectrum (with the resolving power R $\sim 20,000$), using a spectrograph mounted on the $1.82$~m Copernico telescope located at Cima Ekar (Asiago, Italy). Because only a few of our detected candidates are observable from the Asiago observatory, we selected the best observable CEMP candidate, whose \Feh\ was determined by \TC\ to be $-0.96$. The selected star, with V = $12.79$, was on the dark limit of the used telescope, therefore low SNR was expected. The one-hour long exposure of the selected object was fully reduced, normalised order by order, and shifted to the rest frame.

Although the acquired spectrum covers a much wider and continuous spectral range (from 3900 to 7200~\AA) than the HERMES spectra, only subsets, relevant for the classification of carbon-enhanced stars are presented in Figure \ref{fig:asiago}. They were identified by visually matching our observed spectrum with the published moderate-resolution spectral atlas \citep{1996ApJS..105..419B} of peculiar carbon stars. Where available, the GALAH spectrum is shown alongside the Asiago spectrum. Carbon enhancement is not expected to vary over a period of several years, therefore both spectra should show similar features. The second and fourth panel in Figure \ref{fig:asiago} confirm that both observations indicate a similar degree of carbon enhancement.

Following the classification criteria of carbon stars, we determined that the star belongs to the C-H sub-class. The definitive features for this class are strong molecular CH bands, prominent secondary P-branch head near 4342~\AA\ (top panel in Figure \ref{fig:asiago}), and noticeable Ba II lines at 4554 and 6496~\AA\ \citep{2018ApJS..234...31L}, which are all present in the spectrum. The star definitely does not have a high ratio between $^{13}$C and $^{12}$C isotopes as the Swan features corresponding to $^{13}$C are clearly not present, therefore it can not be of a C-J sub-class.

Following the current state of knowledge \citep{1990ApJ...352..709M, 2016AA...586A.158J, 2016ApJ...826...85S} that most, if not all, C-H stars show clear evidence for binarity, we compared the radial velocity between both observations. They hint at the variability of the object as the follow-up radial velocity ($126.75 \pm 1.63$ \kms) deviates by more than $3$~\kms\ from the velocity ($123.43 \pm 0.08$ \kms) observed as part of the GALAH survey. The time span between the two observations is more than 2.5~years, where the exact JD of the observation is $2458090.702$ for the Asiago spectrum, and $2457122.095$ for the GALAH spectrum. Further observations along the variability period would be needed to confirm whether it is a multiple stellar system.

\section{Conclusions}
\label{sec:summary}
This work explores stellar spectra acquired by the HERMES spectrograph in order to discover peculiar carbon-enhanced stars, which were observed in the scope of multiple observing programmes conducted with the same spectrograph.

We show that the spectra of such stars are sufficiently different from other stellar types to be recognisable in high-resolution spectra with limited wavelength ranges. This can be done using a supervised procedure, where some knowledge about the effects of carbon enhancement on the observed spectra is put into the algorithm, or using an unsupervised method. The latter was used to identify observed stars solely on the basis of acquired spectra. By combining both methodologies we identified 918 unique stars with evident signs of carbon enhancement of which 12 were already reported in the literature. Out of all matched objects from the literature, we were unable to detect and confirm 16 ($57$~\%) CH and 15 ($93$~\%) CEMP stars with our procedures. As some of those objects were proven to contain carbon enhancement detectable outside the HERMES wavelength ranges, this would have to be taken into account to say more about the underlying population of carbon-enhanced stars. In addition to a detection bias imposed by the analysis of C$_2$ bands and exclusion of CN, and CH molecular bands that might be excitated in different temperature ranges, varying degree of carbon-enhancement also has to be accounted for accurate population studies. As shown by \citet{2016ApJ...833...20Y}, CEMP stars can be found within a wide range of absolute carbon abundances. When an object selection is performed with a pre-defined threshold, as in the case of our supervised methodology, this may reduce the number of objects in only one of the sub-classes. In the case of CEMP stars, this selection may influence a number CEMP-no stars that are known to have lower absolute carbon abundance \citep{2016ApJ...833...20Y}.

The identified objects were separated into dwarf and giant populations using their stellar atmospheric parameters that were also used to select possible CEMP candidates. All of the detections, with multiple observations at different epochs, were investigated for signs of variability. More than half of the repeats show signs of variability in their measured radial velocities. This could be an indicator that we are looking at a pulsating object or a multiple stellar system.

With a follow-up observation of one of the identified stars, we were able to confirm the existence of carbon-rich molecules in its atmosphere in a wider wavelength range. The acquired spectrum was also used to determine its sub-class. Variation in radial velocity points to a possible variable nature of the star or binarity that is common for C-H stars.

Follow-up observations are required to confirm variability of radial velocities observed for some of the detected carbon-enhanced stars and further investigate their nature. Careful spectral analysis, with the inclusion of carbon enhancement in models, is needed to confirm the metallicity levels of the metal-poor candidates. 

The list of detected stars presented in this paper is accessible as electronic table through the CDS. Detailed structure is presented in Table \ref{tab:out_table}. The list also includes stars from the literature, matched with our observations, for which we were unable to confirm their carbon enhancement. The list could be used to plan further observations, allowing a better understanding of these objects.

\section*{Acknowledgments}
This work is based on data acquired through the Australian Astronomical Observatory, under programmes: A/2013B/13 (The GALAH pilot survey); A/2014A/25, A/2015A/19, A2017A/18 (The GALAH survey); A/2015A/03, A/2015B/19, A/2016A/22, A/2016B/12, A/2017A/14 (The K2-HERMES K2-follow-up programme); A/2016B/10 (The TESS-HERMES programme); A/2015B/01 (Accurate physical parameters of Kepler K2 planet search targets); S/2015A/012 (Planets in clusters with K2). We acknowledge the traditional owners of the land on which the AAT stands, the Gamilaraay people, and pay our respects to elders past and present.

K\v{C}, TZ, and GT acknowledge financial support of the Slovenian Research Agency (research core funding No. P1-0188 and project N1-0040). JK is supported by a Discovery Project grant from the Australian Research Council (DP150104667) awarded to J. Bland-Hawthorn and T. Bedding. DMN was supported by the Allan C. and Dorothy H. Davis Fellowship. SLM acknowledges support from the Australian Research Council through grant DP180101791. Parts of this research were conducted by the Australian Research Council Centre of Excellence for All Sky Astrophysics in Three Dimensions (ASTRO 3D), through project number CE170100013. KF is grateful for support from Australian Research Council grant DP160103747. DS is the recipient of an ARC Future Fellowship (project number FT140100147). Parts of this research were conducted by the Australian Research Council  Centre  of  Excellence  for  All  Sky  Astrophysics  in  three Dimensions (ASTRO 3D), through project number CE170100013. Follow-up observations were collected at the Copernico telescope (Asiago, Italy) of the INAF - Osservatorio Astronomico di Padova.

This work has made use of data from the European Space Agency (ESA) mission {\it Gaia} (\url{https://www.cosmos.esa.int/gaia}), processed by the {\it Gaia} Data Processing and Analysis Consortium (DPAC, \url{https://www.cosmos.esa.int/web/gaia/dpac/consortium}). Funding for the DPAC has been provided by national institutions, in particular the institutions participating in the {\it Gaia} Multilateral Agreement.

\bibliographystyle{mnras}
\bibliography{bib}

\appendix
\section{Table description} 
In the Table \ref{tab:out_table} we provide a list of metadata available for every object detected using the methodology described in this paper. The complete table of detected objects and its metadata is available only in electronic form at the CDS.

\begin{table}
\caption{List and description of the fields in the published catalogue of detected objects and objects matched with multiple literature sources.}
\label{tab:out_table}
\begin{tabular}{l c l}
\hline
Field & Unit & Description \\ 
\hline
\texttt{source\_id} & & {\it Gaia} DR2 source identifier \\
\texttt{sobject\_id} & & Unique internal per-observation star ID \\
\texttt{ra} & deg & Right ascension from 2MASS, J2000 \\
\texttt{dec} & deg & Declination from 2MASS, J2000 \\
\texttt{det\_sup} & bool & Detected by supervised fitting method \\
\texttt{det\_usup} & bool & Detected by t-SNE method \\
\texttt{swan\_integ} & & Swan band strength if determined \\
\texttt{teff} & K & \TC\ effective temperature \Teff \\
\texttt{e\_teff} & K & Uncertainty of determined \Teff \\
\texttt{logg} & & \TC\ surface gravity \Logg \\
\texttt{e\_logg} & & Uncertainty of determined \Logg \\
\texttt{feh} & & \TC\ iron abundance \Feh \\
\texttt{e\_feh} & & Uncertainty of determined \Feh \\
\texttt{flag\_cannon} & int & \TC\ flags in a bit mask format \\
\texttt{type} &  & G for giants and D for dwarfs \\
\texttt{rv\_var} & bool & Is radial velocity variable \\
\texttt{li\_strong} & bool & Shows strong lithium absorption \\
\texttt{cemp\_cand} & bool & Is star CEMP candidate \\
\texttt{bib\_code} &  & ADS bibcode of the literature match \\
\hline
\end{tabular}
\end{table}

\begin{figure*}
	\centering
	\includegraphics[width=0.95\textwidth]{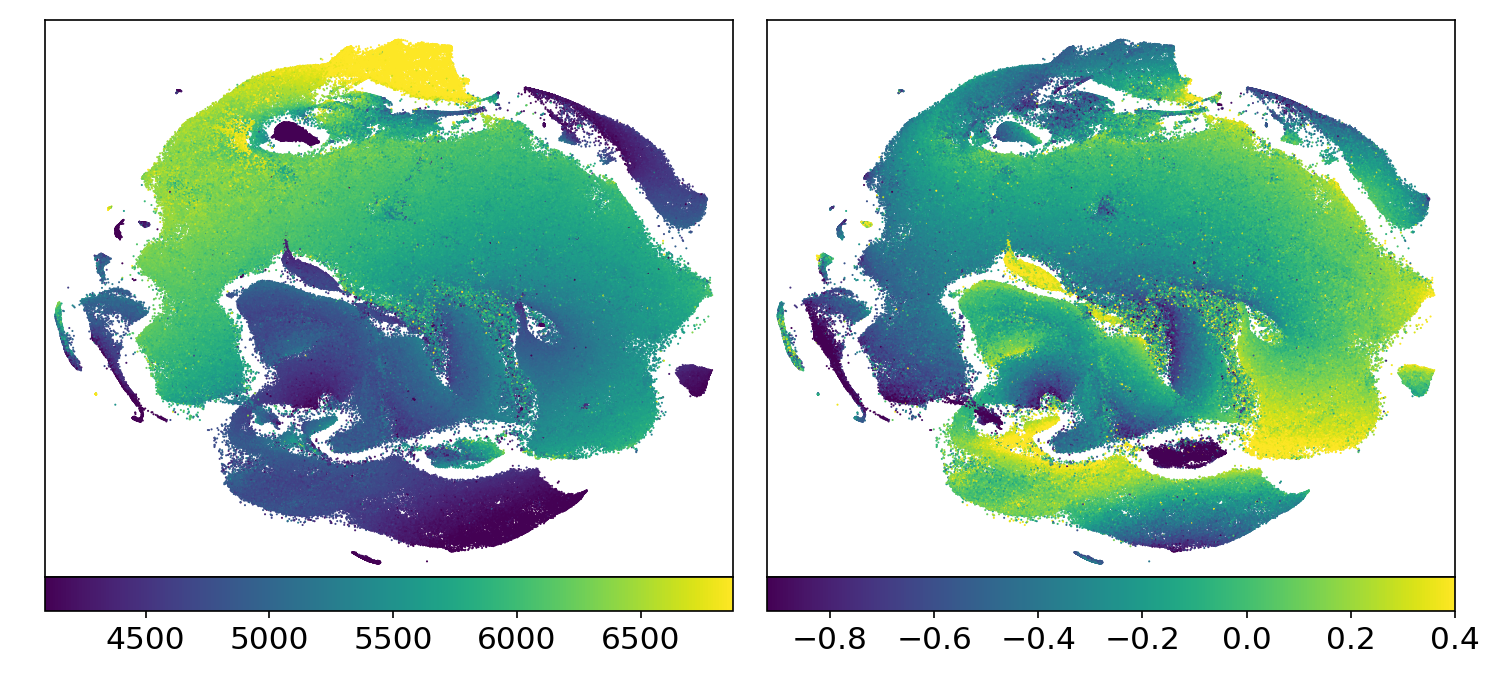}
	\caption{Spatial distribution of all available measurements of \Teff\ (left panel) and \Feh\ (right panel) as determined by \TC. Dots, representing analysed spectra in the t-SNE projection, are colour coded by their parameter values. Colours and their corresponding values are explained by a colourbar under the graph.}
	\label{fig:tsne_teff_feh}
\end{figure*}

\begin{figure*}
	\centering
	\includegraphics[width=0.95\textwidth]{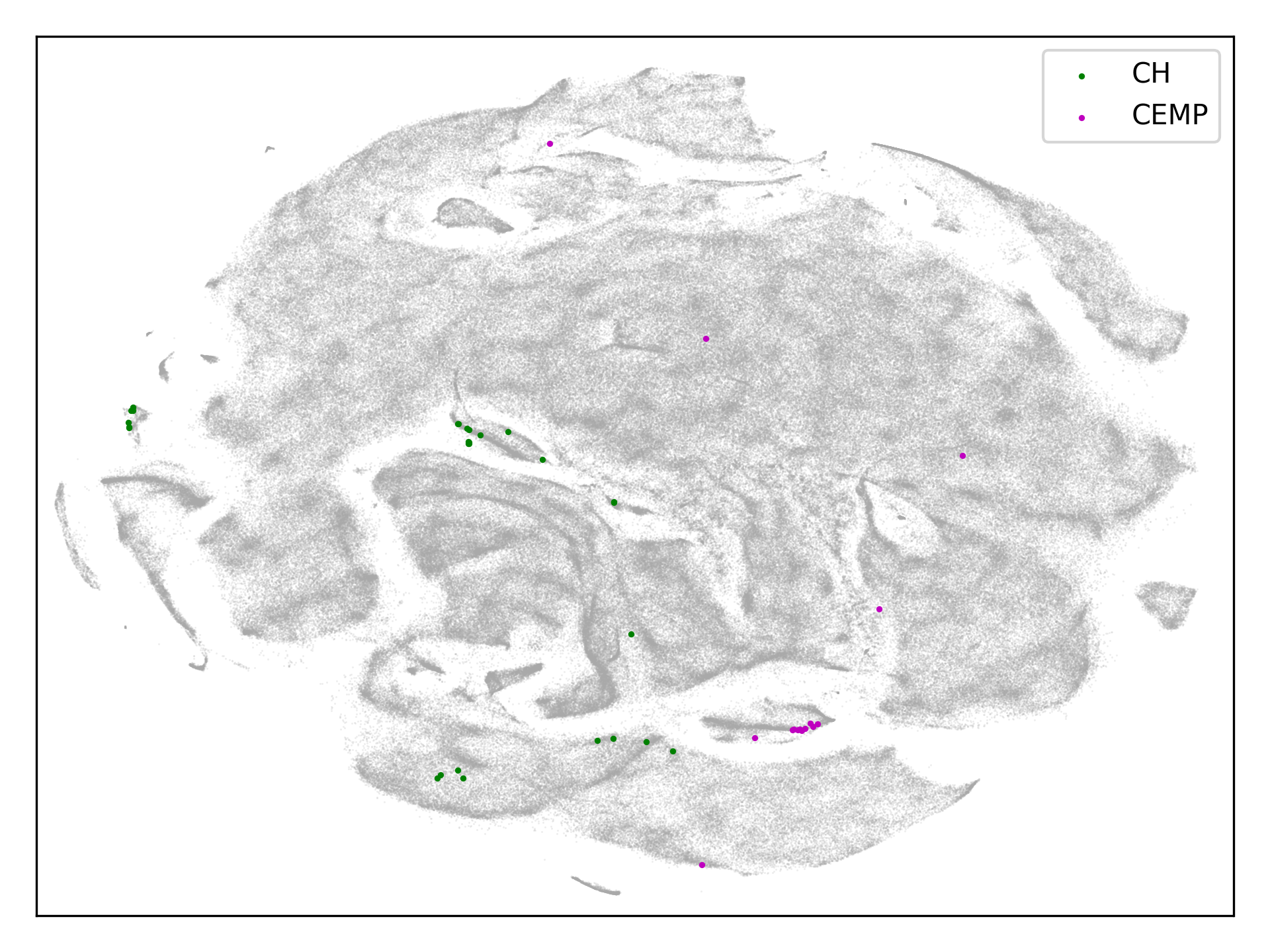}
	\caption{t-SNE projection with marked known carbon-enhanced and CEMP objects from multiple different catalogues found in the literature that are also part of our analysed set of spectra.}
	\label{fig:tsne_ref_ch}
\end{figure*}

\begin{figure*}
	\centering
	\includegraphics[width=\textwidth]{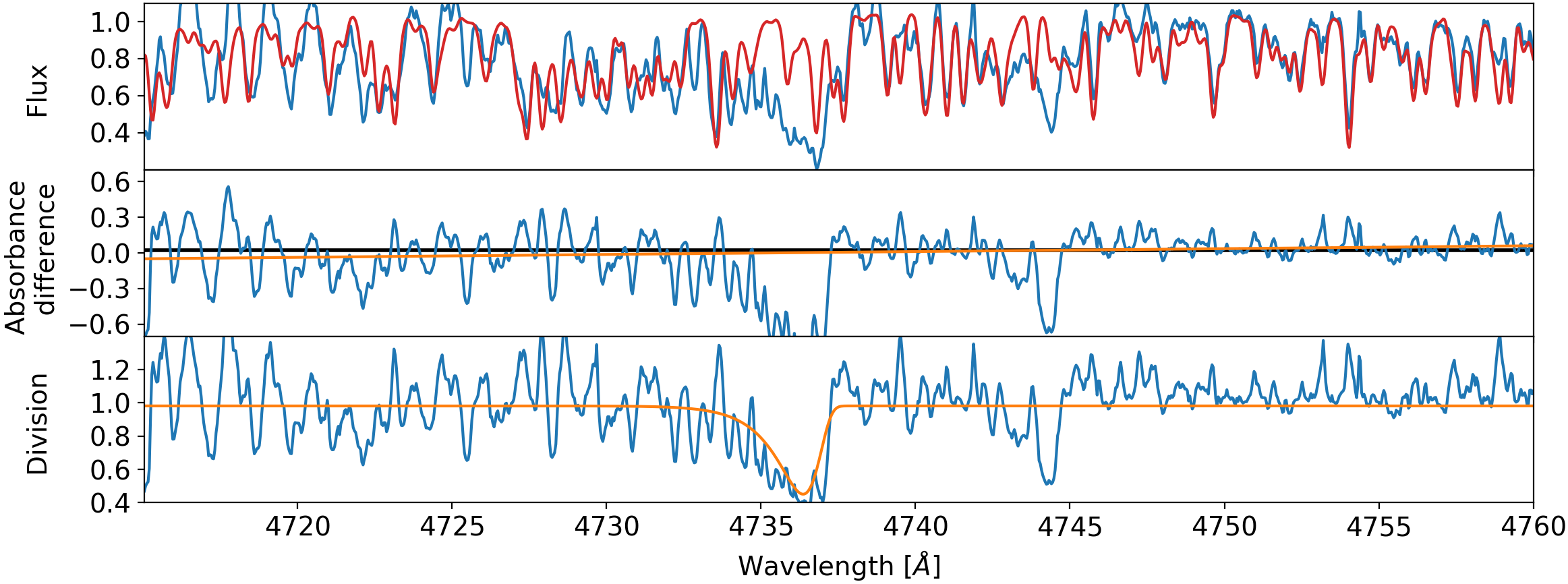}
	\caption{Equivalent plot as in the Figure \ref{fig:carbon_example} but presenting an example of a metal-rich star with multiple strong Swan features around 4737 and 4745 \AA. Presented star has a 2MASS identifier J13121354-3533120 and is known Galactic carbon star \citep{2001BaltA..10....1A}.}
	\label{fig:carbon_example2}
\end{figure*}

\begin{figure*}
	\centering
	\includegraphics[width=\textwidth]{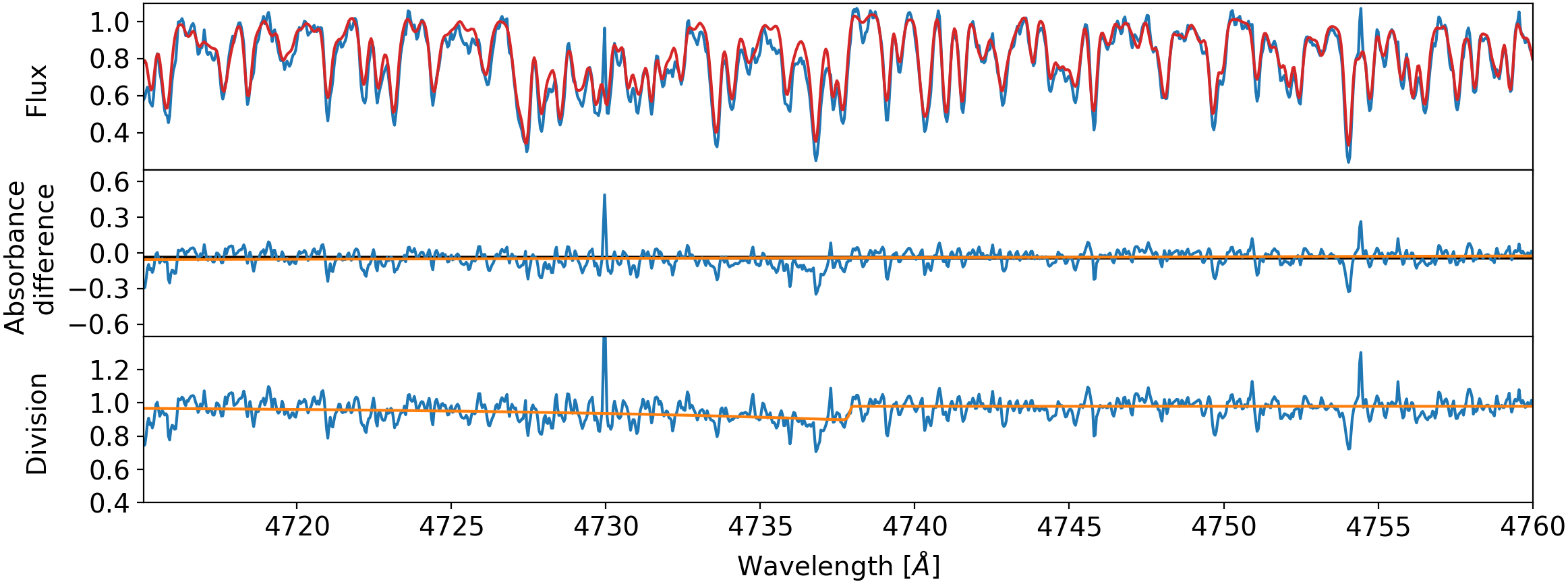}
	\caption{Equivalent plot as in the Figure \ref{fig:carbon_example} showing the last of 400 spectra, ordered by their degree of carbon enhancement, selected by the supervised methodology.}
	\label{fig:carbon_last_supervised}
\end{figure*}

\newpage

\begin{figure*}
	\centering
	\includegraphics[width=\textwidth]{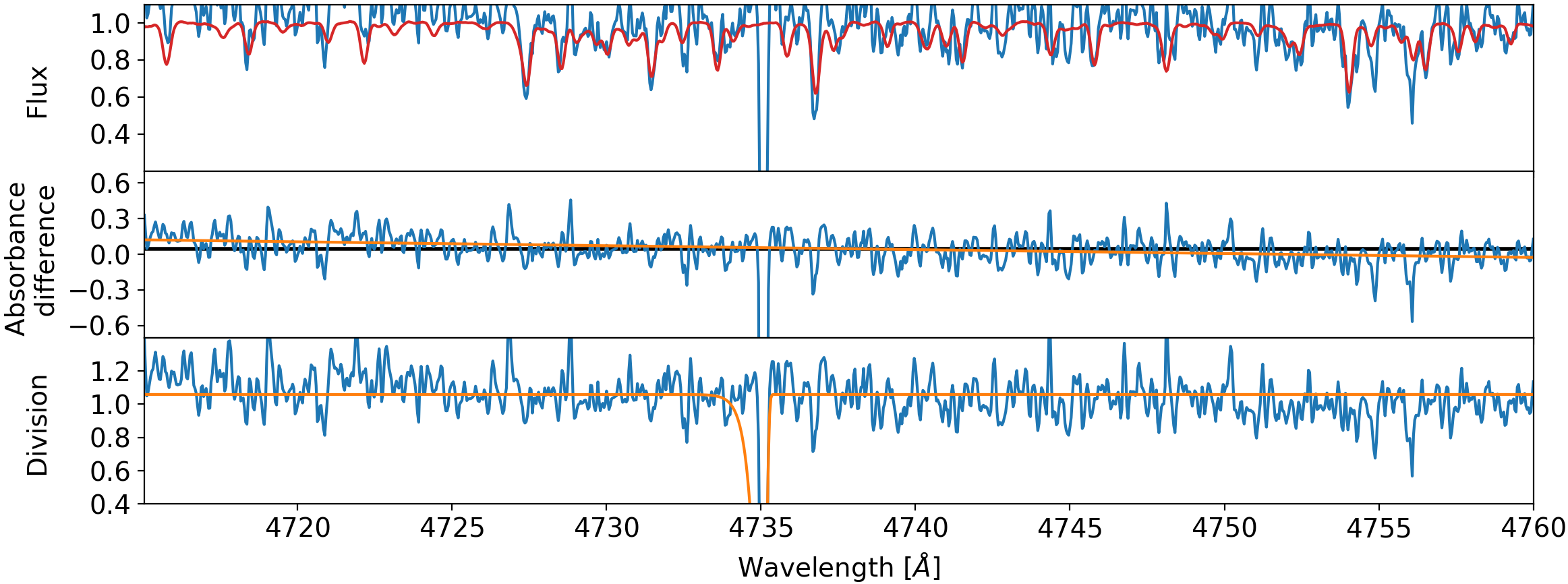}
	\caption{Equivalent plot as in the Figure \ref{fig:carbon_example} but representing grossly over exaggerated carbon enhancement by a fit that describes a reduction problem (a cosmic ray in a subtracted sky spectrum).}
	\label{fig:bad_fit1}
\end{figure*}

\begin{figure*}
	\centering
	\includegraphics[width=\textwidth]{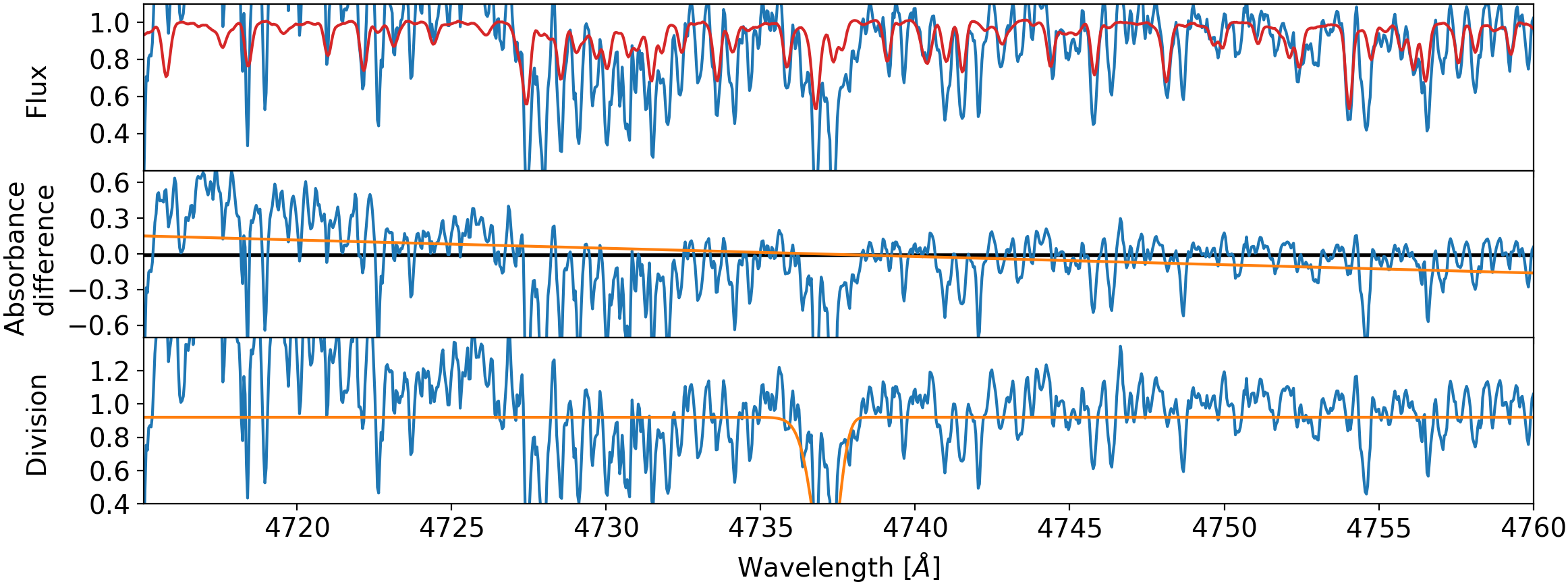}
	\caption{Equivalent plot as in the Figure \ref{fig:carbon_example} but representing a fit to absorption lines of a double-lined spectroscopic binary. Final fit is not skewed as would be expected in the case of carbon enhancement.}
	\label{fig:bad_fit2}
\end{figure*}

\begin{figure*}
	\centering
	\includegraphics[width=\textwidth]{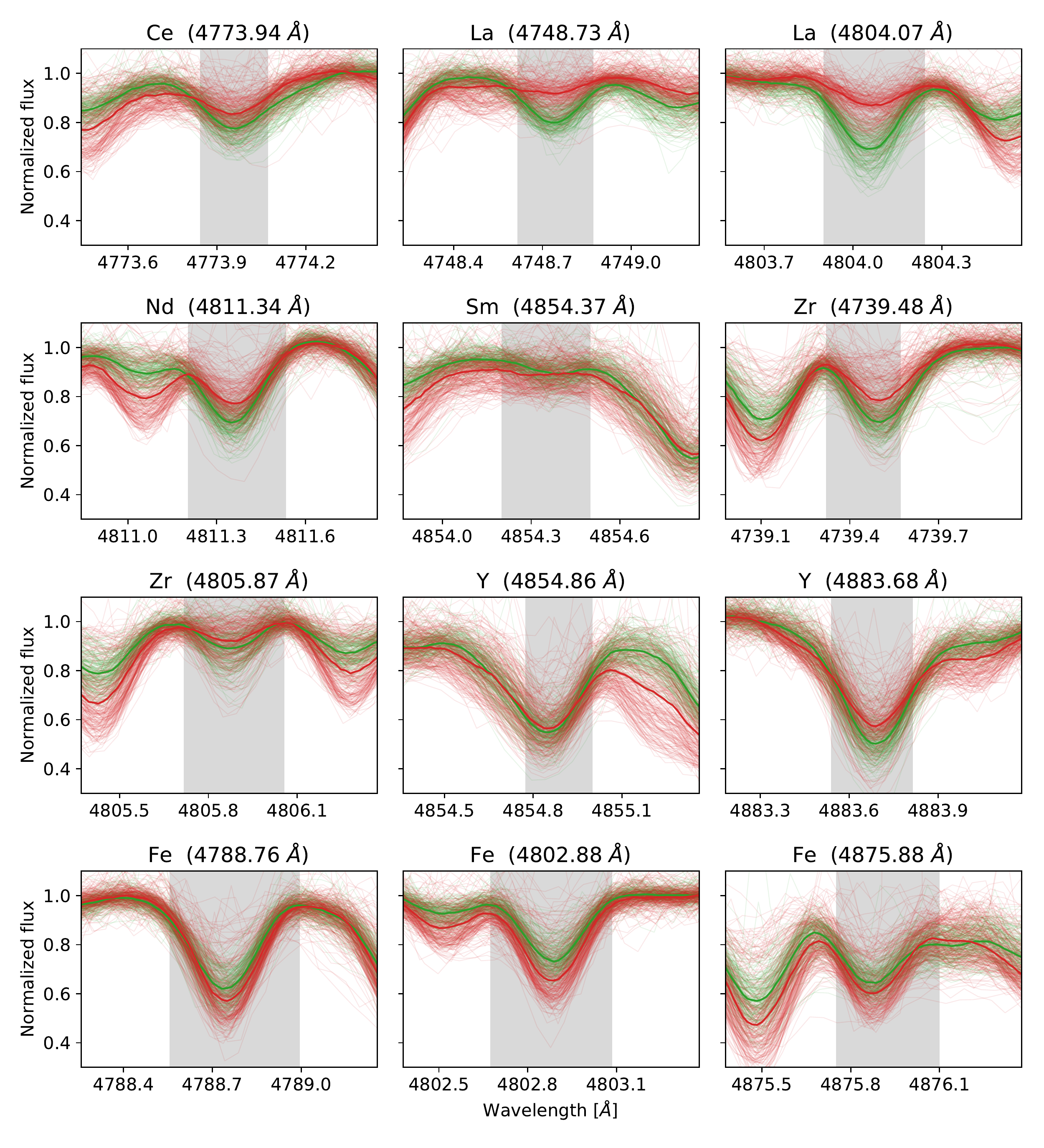}
	\caption{Spectral subset around the absorption features in the blue arm that were used to determine abundances of Fe and s-process elements. Same colour coding is used as in Figure \ref{fig:sprocess_hist}. Spectra inside the t-SNE determined clump are shown in red, and outside it in green. Median of all spectra is shown with a bold line of the same colour. The shaded area gives the wavelength range considered in the computation of abundances.}
	\label{fig:sprocess_spectrum}
\end{figure*}

\begin{figure}
	\centering
	\includegraphics[width=\columnwidth]{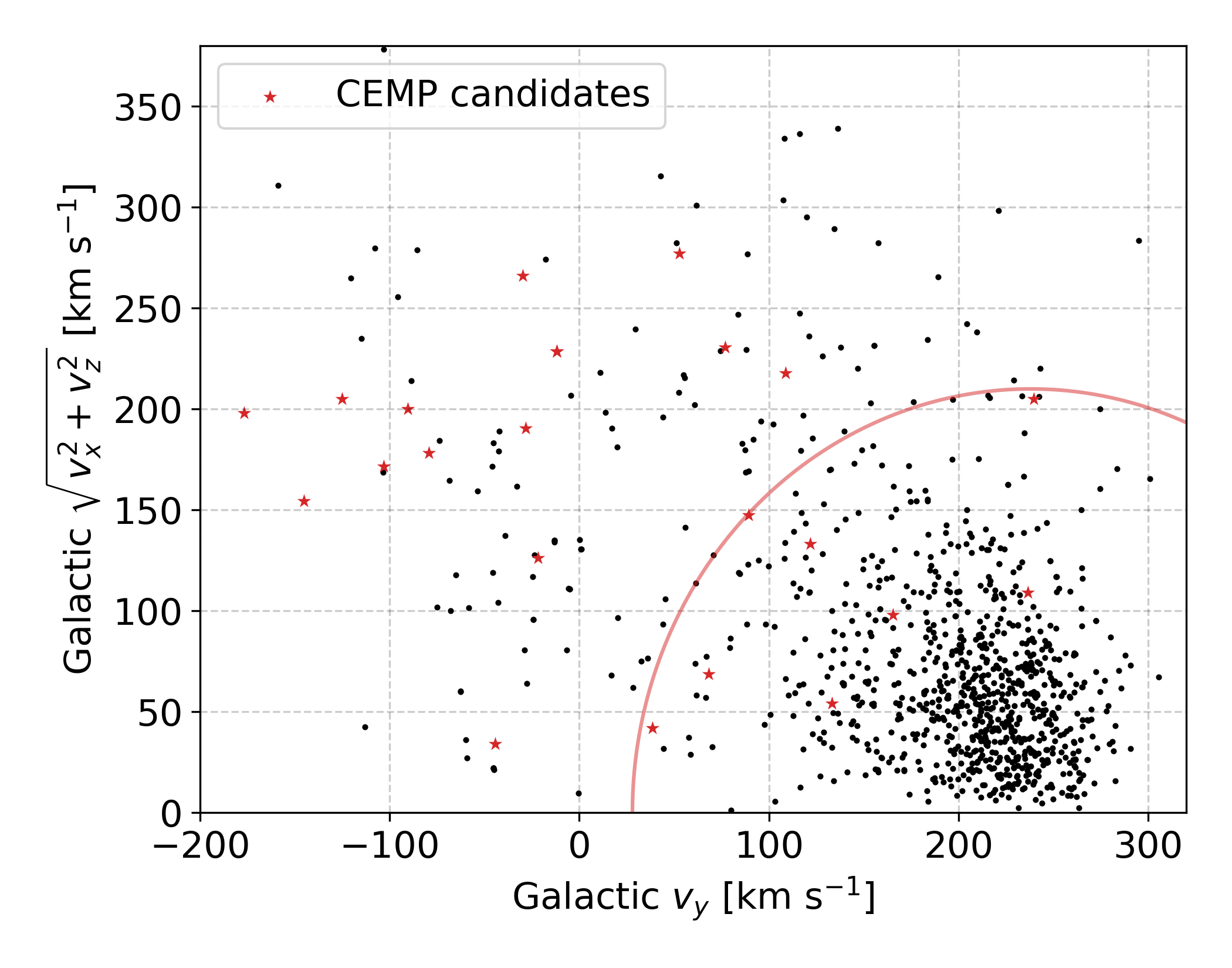}
	\caption{Toomre diagram used to identify possible local halo stars among our detected carbon-enhanced stars, especially CEMP candidates. Halo stars in this diagram are located above the red circular line, satisfying the velocity condition $\left|\mathbf{v} - \mathbf{v_{LSR}} \right|$~>~210~\kms\ \citep[the threshold taken from ][]{2018ApJ...860L..11K}. CEMP candidates are marked with star symbols.}
	\label{fig:orbits_vxvyvz}
\end{figure}

\bsp	
\label{lastpage}
\end{document}